\documentclass[12pt,a4paper]{article}
\usepackage{amsmath}
\usepackage{amssymb}
\usepackage{epsfig}
\usepackage{subfigure}
\newcommand{\V}[1]{V_{#1}^{\phantom{\ast}}}
\newcommand{\Vc}[1]{V_{#1}^{\ast}}
\newcommand{\abs}[1]{\left|#1\right|}
\newcommand{\absV}[1]{\left|V_{#1}\right|}

\begin{document}

\vspace*{-3cm}
\begin{flushright}
hep-ph/0502133 \\
CERN-PH-TH/2005-015\\
IFIC/05-11\\
FTUV-05-0214\\
February 2005
\end{flushright}

\begin{center}
\begin{Large}
{\bf New Physics and Evidence\\ for a Complex CKM }
\end{Large}
 
\vspace{0.5cm}
F.J. Botella $^a$, G.C. Branco $^b$, M. Nebot $^a$ and M.N. 
Rebelo $^b$\\[0.2cm]
{\it $^a$ Departament de F\'{\i}sica Te\`orica and IFIC,\\
Universitat de Val\`encia-CSIC,\\ E-46100, Burjassot, Spain.}\\
{\it $^b$ Physics Department, Theory, CERN,\\ CH-1211 Geneva 23, Switzerland.}
\footnote{On leave of absence from Departamento de F\'{\i}sica and
Centro de F\'{\i}sica Te\'orica de Part\'{\i}culas (CFTP),
Instituto Superior T\'ecnico, Av. Rovisco Pais, 1049-001 Lisboa,\\ Portugal.}
 \end{center}

\begin{abstract}
We carefully analyse the present experimental evidence 
for a complex CKM matrix, even allowing for New Physics contributions
to  $\epsilon _{K}$, $a_{J/\Psi K_S}$, $\Delta M_{B_{d}}$, 
$\Delta M_{B_{s}}$, and the $\Delta I=1/2$ piece of $B\to\rho\rho$ and $B\to\rho\pi$. We emphasize the crucial r\^ ole played by the
angle $\gamma$ in both providing irrefutable evidence for a complex 
CKM matrix and placing constraints on the size of NP contributions.
It is shown that even if one allows for New Physics a real CKM matrix is excluded at a 99.92\% C.L., and the probability for the phase $\gamma$ to be in the interval $[-170^\circ;-10^\circ]\cup [10^\circ;170^\circ]$ is 99.7\%.
\end{abstract}

\section{Introduction}

At present, the Cabibbo--Kobayashi--Maskawa (CKM) 
\cite{Cabibbo:1963yz}
mechanism for flavour mixing
and CP violation is in agreement with all available experimental data. This
is a remarkable success, since it is achieved with a relatively small number
of parameters. Once the experimental values of 
$\absV{us},\absV{cb}$ and $\absV{ub}$ are used
to fix the angles $\theta_{12},\theta_{23}$ and $\theta_{13}$ of the
standard parametrization, one has to fit, with a single parameter $\delta
_{13}$, the experimental values of a large number of quantities, including
$\epsilon _{K},\sin \left( 2\beta \right)
,\Delta M_{B_{d}}$, as well as the bound on 
$\Delta M_{B_{s}}$. This impressive result is nicely
represented in the usual unitarity triangle fits 
\cite{Bona:2005vz}. In view of 
the remarkable success of the Standard Model, it is plausible that the
CKM mechanism gives the dominant contribution to mixing and CP violation 
at low energies, 
although there is still significant room for New Physics (NP).

In this paper, we address the question of 
how the data available at present
already provide an irrefutable proof that 
the CKM matrix is ``non-trivially
complex'', thus implying that the charged weak-current interactions violate
CP. Obviously, in the framework of the Standard Model (SM), the CKM matrix
has to be complex, in order to account for the observed CP violation, both
in the kaon and in the B sector. The above question only becomes non-trivial
if one allows for the presence of NP \cite{Branco:1999sx}. 
We shall carefully analyse
the present experimental indications in favour of a complex CKM matrix,
correcting at the same time various misleading and in some cases\ erroneous
statements one finds in the literature.

To have an irrefutable proof that the CKM matrix is complex is of the utmost
importance in order to investigate the origin of CP violation, as well as to
analyse what classes of theories of CP violation are viable in view of the
present experimental data provided by B-factories. For example, one of the
crucial questions concerning the origin of CP violation is whether CP is
violated explicitly in the Lagrangian through the introduction of 
complex Yukawa
couplings or, on the contrary, CP is a good symmetry of the Lagrangian, only
spontaneously broken by the vacuum. There are two classes of theories with
spontaneous CP violation:

\begin{enumerate}
\item[i)] Those where the phases arising from the vacuum lead to a complex
CKM matrix, in spite of having real Yukawa couplings. In this class of
theories, there are in general more than one source of CP violation, namely
the usual KM mechanism, together with some new sources of CP violation.

\item[ii)] Models where all phases can be removed from 
the CKM matrix and thus
CP violation arises exclusively from New Physics.
\end{enumerate}

Examples of  class i) are, for instance, two Higgs doublet extensions
of the SM with spontaneous CP violation, without natural
flavour conservation (NFC) in the Higgs sector \cite{Lee:1973iz,Branco:1985aq},
as well as models where CP is broken at a high energy scale
\cite{Bento:1990wv} 
through a phase in the vacuum expectation value (vev)
of a complex singlet, which in turn generates a non-trivially complex 
CKM matrix through the mixing of isosinglet quarks with standard 
quarks. Examples of class ii) are the simplest supersymmetric
extensions of the SM with spontaneous CP violation  \cite{Branco:2000dq},
as well as three Higgs doublet models with spontaneous CP
violation \cite{Branco:1979pv} 
  and NFC in the Higgs sector.

It is clear that if it can be proved that experimental data constrain the CKM
matrix to be complex, even allowing for the presence of New Physics, the only
models of spontaneous CP violation that are viable belong to class i).

The paper is organized as follows. In the next section 
we specify what our assumptions about NP are and we address 
the question of how to obtain evidence for a complex CKM matrix 
which would not be affected by the presence of this type of NP.
In section 3 we discuss the 
possibility of obtaining an experimental proof that the CKM matrix
is complex, from the measurement of four indepedent moduli
of this matrix. We argue that the cleanest proof would be
obtained if one would use only moduli from the first two rows
of the CKM matrix. However, we point out that contrary to
some statements in the literature, this proof cannot be
obtained in practice, since it would require totally unrealistic
precision in the measurement of  $\absV{us}$,  $\absV{cd}$.
In section 4 we analyse the impact of the measurement of 
$\Delta M_{B_{d}}$ and $a_{J/\Psi K_S}$ and introduce a 
convenient parametrization of NP. In section 5 we emphasize 
the importance of a measurement of $\gamma$ either direct or
through the combined measurement of $\overline \beta$,
$\overline \alpha$. Finally, we present 
our summary and conclusions in section 6.

\section{Evidence for a complex CKM matrix unaffected by the 
presence of New Physics}

Our assumptions about NP will be the following. We will assume that in weak
processes where the SM contributes at tree level, NP contributions are
negligible, while NP may be relevant to weak processes where the SM only
contributes at the loop level. This is a rather mild assumption on NP.
In addition, for the moment we will also
assume that the 3$\times$3 CKM matrix is unitary, 
but we will come back to this point
later. An example of this framework is the minimal supersymmetric standard
model (MSSM).

This scenario implies that NP could be competing 
with any SM loop and more
likely in $K^{0}$--$\bar{K}^{0}$, $B_{d}^{0}$--$\bar{B}_{d}^{0}$,
$B_{s}^{0}$--$\bar{B}_{s}^{0}$ 
and $D^{0}$--$\bar{D}^{0}$ mixing. Therefore, the
usual interpretation of $\epsilon _{K},a_{J/\Psi K_S},\Delta M_{B_{d}},\Delta
M_{B_{s}}$ in terms of $\delta _{13}$ gets invalidated. The same  could also be
the case for the direct CP-violating parameters 
$\epsilon ^{\prime }/\epsilon $, 
$A_\mathrm{dir}^\mathrm{CP}(B^0\to K^+\pi^-)$ \cite{Aubert:2004qm} 
and various semileptonic rare decays. To clarify which are the
fundamental observables that will not be polluted by 
NP contributions, let us
write the CKM matrix in a particular phase convention  
\cite{Branco:1999fs} in terms of all the
rephasing invariant quantities \cite{Aleksan:1994if,Botella:2002fr}
that can a priori be measured in the flavour sector:
\begin{equation}
V_\mathrm{CKM}=\left( 
\begin{array}{ccc}
\absV{ud} & \absV{us} e^{i\chi^{\prime }} &\absV{ub} e^{-i\gamma } \\ 
-\absV{cd} & \absV{cs} & \absV{cb} \\ 
\absV{td} e^{-i\beta } & -\absV{ts}e^{i\chi } & \absV{tb}
\end{array}%
\right)  \label{VCKM1}
\end{equation}%
where the CP-violating phases introduced in Eq. (\ref{VCKM1}) 
are defined by:
\begin{equation}
\begin{array}{ccc}
\beta =\arg \left( -\V{cd}\Vc{cb}\Vc{td}\V{tb}\right) & , & 
\gamma =\arg \left( -\V{ud}\Vc{ub}\Vc{cd}\V{cb}\right)~, \\ 
\chi =\arg \left( -\V{ts}\Vc{tb}\Vc{cs}\Vc{cb}\right) & , & 
\chi ^{\prime }=\arg \left( -\V{cd}\Vc{cs}\Vc{ud}\V{us}\right)~.
\end{array}
\label{CPphas0}
\end{equation}%
Note that $\alpha \equiv \arg \left( -\V{td}\Vc{tb}\Vc{ud}\V{ub}\right)$
obeys the relation  
$ \alpha =\pi -\beta -\gamma $, by definition. Without imposing the
constraints of unitarity, the four rephasing invariant phases, 
together with the nine moduli are all the independent physical
quantities contained in $V_\mathrm{CKM}$. In the SM, where unitarity
holds, these quantities are related by a series of exact relations 
which provide a stringent test of the SM  \cite{Botella:2002fr,SilvaWolf}.

It is clear that the moduli of the first two rows are extracted from weak
processes at tree level in the SM. From top decays we have some 
direct information
on $\absV{tb}$ -- useless in practice compared to its 
unitarity determination from $\absV{ub}$ and 
$\absV{cb}$ --, and essentially no direct information on $\absV{ts} $
and $\absV{td}$ from tree level decays. Therefore, the extraction of the moduli
of the third row is made with loop processes and consequently can be
affected by the presence of NP. From Eq. (\ref{VCKM1}) it is evident that the
phases $\beta $ and $\chi $ will only enter in loop amplitudes because they
appear in transitions involving the top quark, and therefore the usual
extraction of these phases could be 
also contaminated by NP. $\chi ^{\prime }$ is
too small to be considered (see references 
\cite{Botella:2002fr,Branco:1999fs}). 
Therefore,  $\gamma $ is the only
phase that can be measured without NP contamination, since it
enters in $b\rightarrow u$ transitions not necessarily involving loop mediated
processes.

In our framework, the most straightforward irrefutable proof of a complex
CKM could come from the knowledge of the moduli of the first two rows of the
CKM matrix and/or from a determination of $\gamma $ in a weak tree level
process. To use the information on $\beta ,\chi $ and the moduli of the
third row will require a much more involved analysis including the presence
of NP. But by the same token, this analysis will be 
extremely interesting in the
not less important goal of discovering the presence of NP. In our search for
an irrefutable proof of complex CKM, we will also analyse in parallel its
consequences for the detection of the presence of NP \cite{Charles:2004jd}, confirming the recent results by Ligeti \cite{Ligeti:2004ak}.

\section{Obtaining Im $\boldsymbol{(V_{i\alpha}V_{j\beta}V_{i\beta}^{\ast}V_{j\alpha}^{\ast})=J}$ from four independent moduli}\label{sec:02}

Within this class of models, in order to investigate whether the present
experimental data already implies that CKM is complex, one has to check
whether any of the unitarity triangles is 
constrained by data to be non-``flat'',
i.e. to have a non-vanishing area. If any one of the triangles does
not collapse to a line, no other triangle will  collapse, 
due to the remarkable property
that all the unitarity triangles have the same area. This property simply
follows from unitarity of the 3$\times$3 CKM matrix. The universal area of the
unitarity triangles gives a measurement of the strength of CP violation
mediated by a $W$-interaction and can be obtained from four independent moduli
of $V_\mathrm{CKM}$. The fact that one can infer about CP violation from the
knowledge of CP-conserving quantities should not come as a surprise 
\cite{Botella:1985gb}. 
It just
reflects the fact that the strength of CP violation is given by the
imaginary part of a rephasing invariant quartet \cite{Jarlskog:1985ht}, 
$J=\pm \text{Im}\left(
\V{i\alpha }\V{j\beta }\Vc{i\beta }\Vc{j\alpha }\right)$, with
$\left( i\neq j,\alpha \neq \beta \right) $, which in turn can be expressed
in terms of moduli, thanks to 
3$\times$3 unitarity. Restricting ourselves to the first
two rows of $V_\mathrm{CKM}$, to avoid any contamination from NP, 
a possible choice
of independent moduli would be $\absV{us}$, $\absV{cb}$, $\absV{ub}$ and 
$\absV{cd}$. One can then use unitarity of the first two rows to
evaluate $J$, which is given, in terms of the input moduli, by
\begin{eqnarray}
4J^{2} &=&4\left( 1-\absV{ub}^{2}-\absV{us}^{2}\right) 
\absV{ub}^{2}\absV{cd}^{2}\absV{cb}^{2}-  \notag \\
&&-\left( \absV{us}^{2}-\absV{cd}^{2}+\absV{cd}^{2}\absV{ub}^{2}
-\absV{cb}^{2}\absV{ub}^{2}-\absV{cb}^{2}\absV{us}^{2}\right)^{2}.~  
\label{Jsquare1}
\end{eqnarray}%
Note that Eq. (\ref{Jsquare1}) is exact, but the actual extraction of $J$
from the chosen input moduli, although possible in principle, it is not
feasible in ``practice''.

\begin{figure}[htb]
\begin{center}
\subfigure[The complete distribution.]
{\epsfig{file=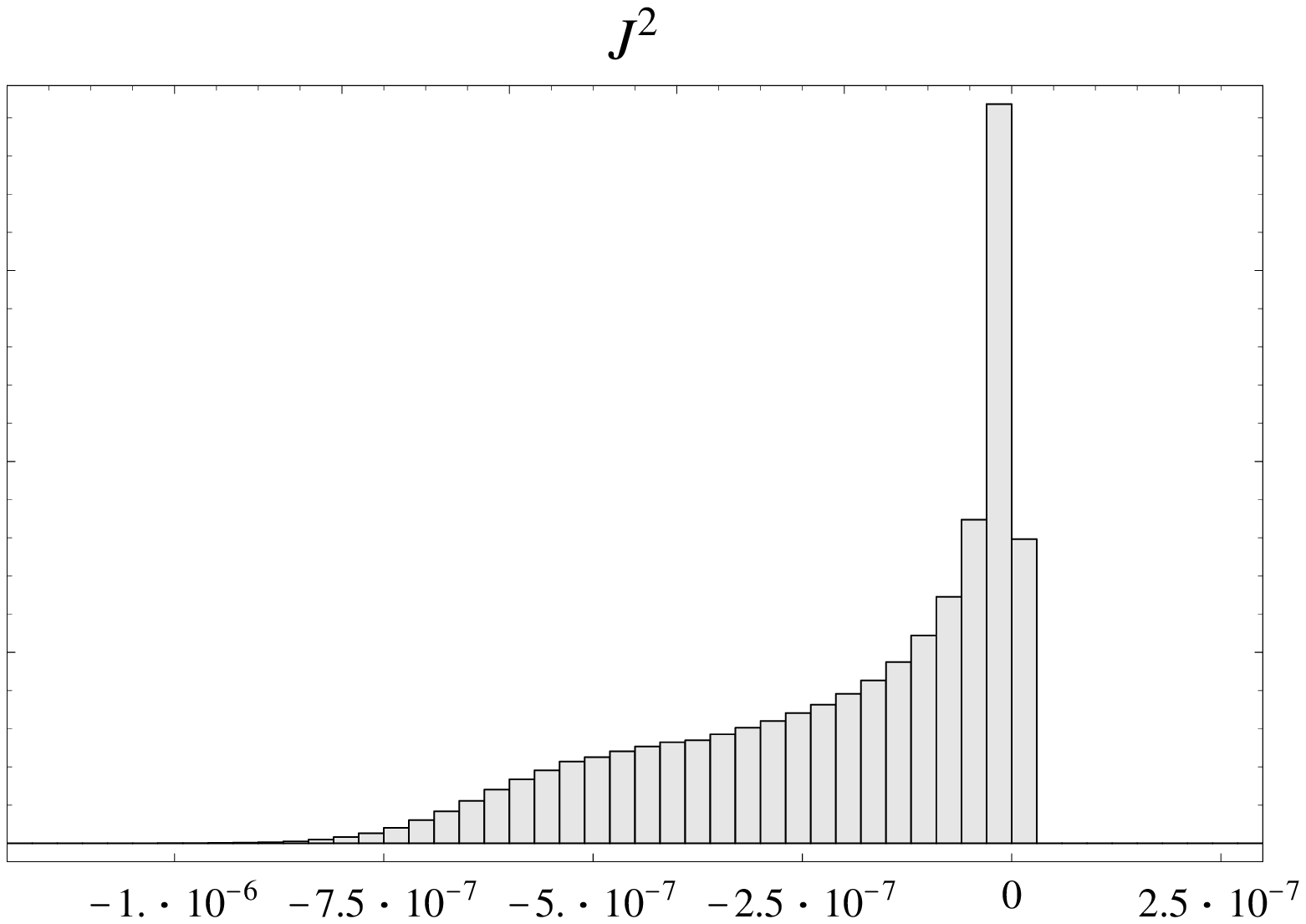,width=0.45\textwidth}}~~\subfigure[The 
region $J^2\sim 0$.]{\epsfig{file=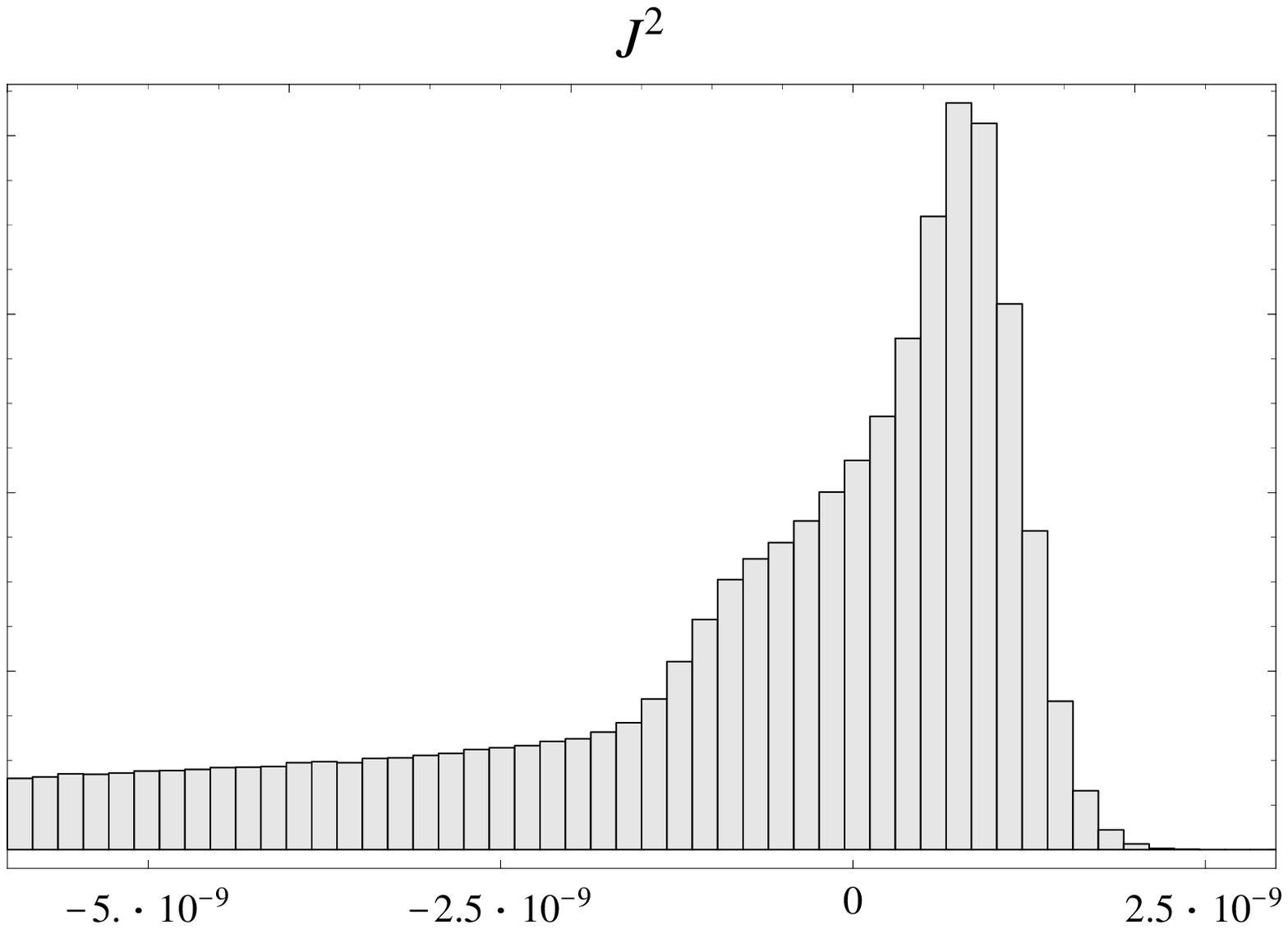,width=0.45\textwidth}}
\end{center}
\caption{$J^{2}$ distribution from $\left| V_{us} \right| 
=0.2200\pm 0.0026 $, 
$\left| V_{cb} \right| = (4.13\pm0.15)10^{-2} $, $\left| V_{ub} \right| 
=(3.67\pm 0.47)10^{-3}  $ and $\left| V_{cd} \right| = 0.224\pm 0.012 $. }
\label{Fig:01}
\end{figure}

To illustrate this point, let us consider the present experimental values of 
$\absV{us},\absV{cb} ,\absV{ub} $ and $\absV{cd} $, assuming Gaussian 
probability density 
distributions around the central values. We plot in Fig. \ref{Fig:01} 
the probability density distribution
of $J^{2}$, generated using a toy Monte Carlo calculation 
\cite{Cowan:1998ji}.
Only $31.1\%$ 
of the generated points satisfy the trivial normalization 
constraints and, among those,
only $7.9\%$ satisfy the condition that the unitarity triangles close 
$\left( J^{2}>0\right) $.

\begin{figure}[htb]
\begin{center}
\epsfig{file=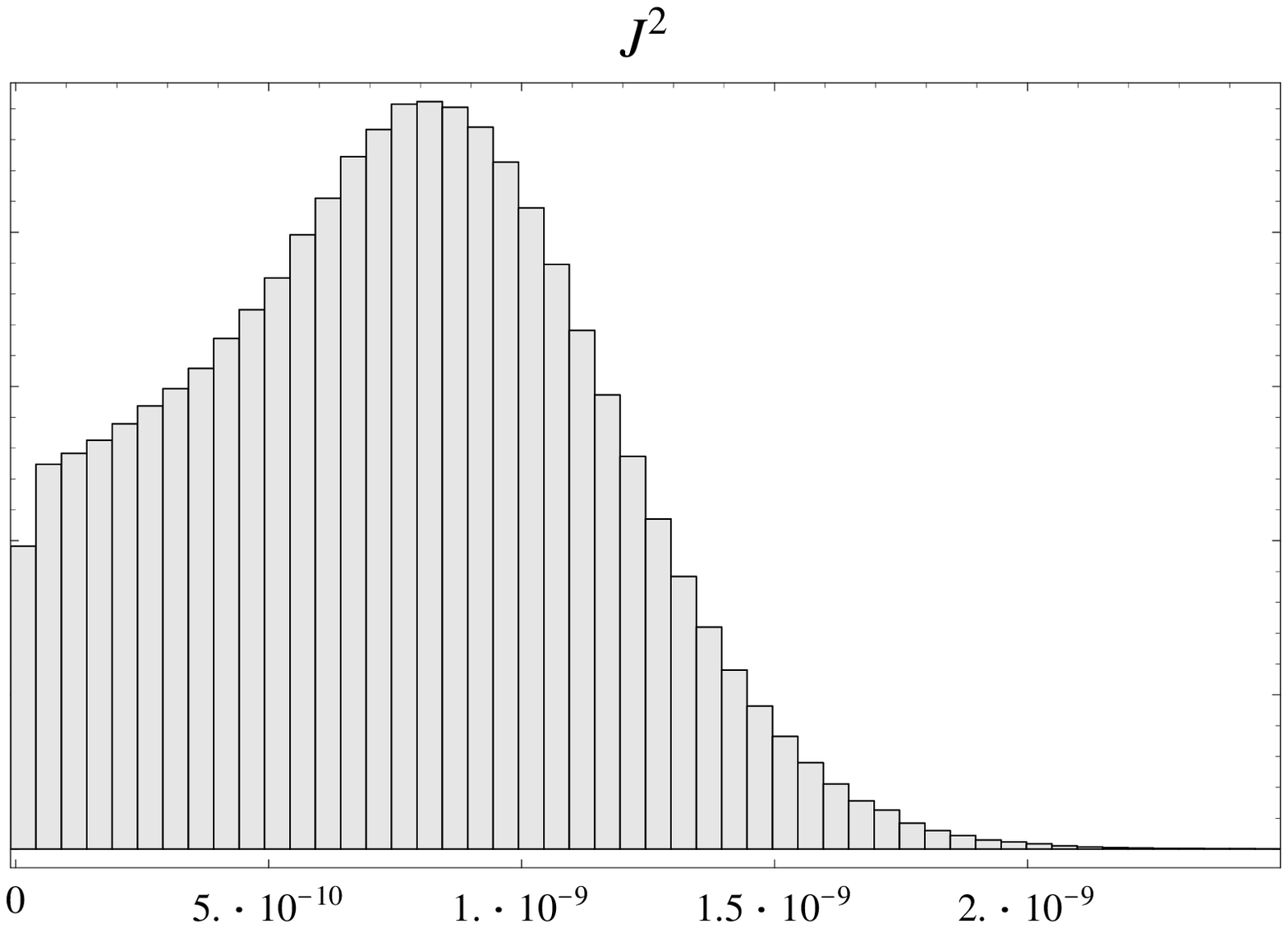,width=0.45\textwidth}~~\epsfig{file=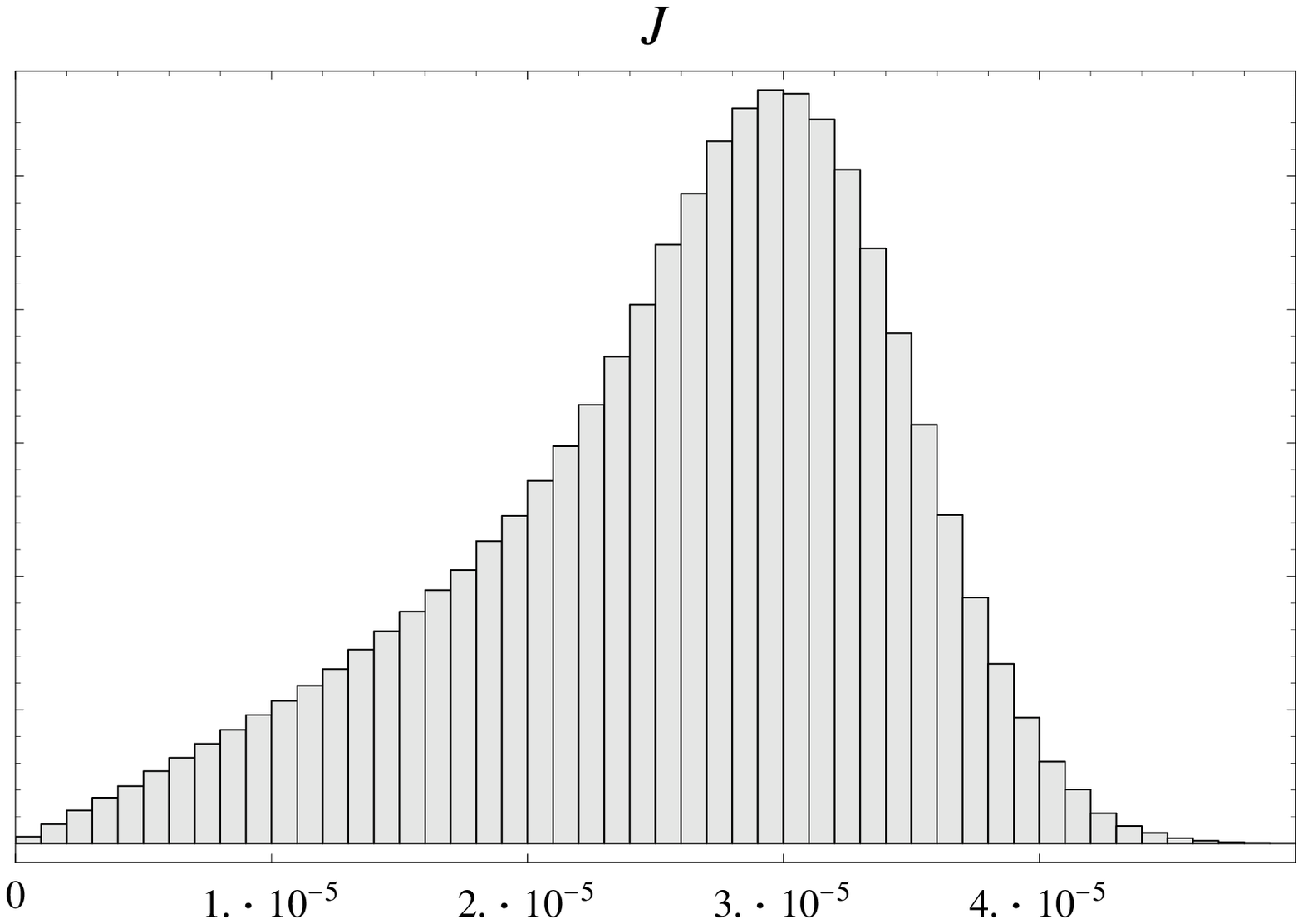,width=0.45\textwidth}
\end{center}
\caption{$J^{2}$ and $J$ distributions.}
\label{Fig:02}
\end{figure}

In order to extract information on $J^{2}$ from the input moduli, it may be
tempting to plot only the points that satisfy all unitarity constraints, i.e.
normalization of columns and rows of $V_\mathrm{CKM}$, 
together with the constraint
of having $J^{2}>0$. We have plotted these points in Fig. \ref{Fig:02}, which naively
lead to $J=\left( 2.6\pm 0.8\right) \times 10^{-5}$.%

\begin{figure}[htb]
\begin{center}
\subfigure[$\absV{cd}=0.21985\pm 0.00002$\label{Fig:03a}]{\epsfig{file=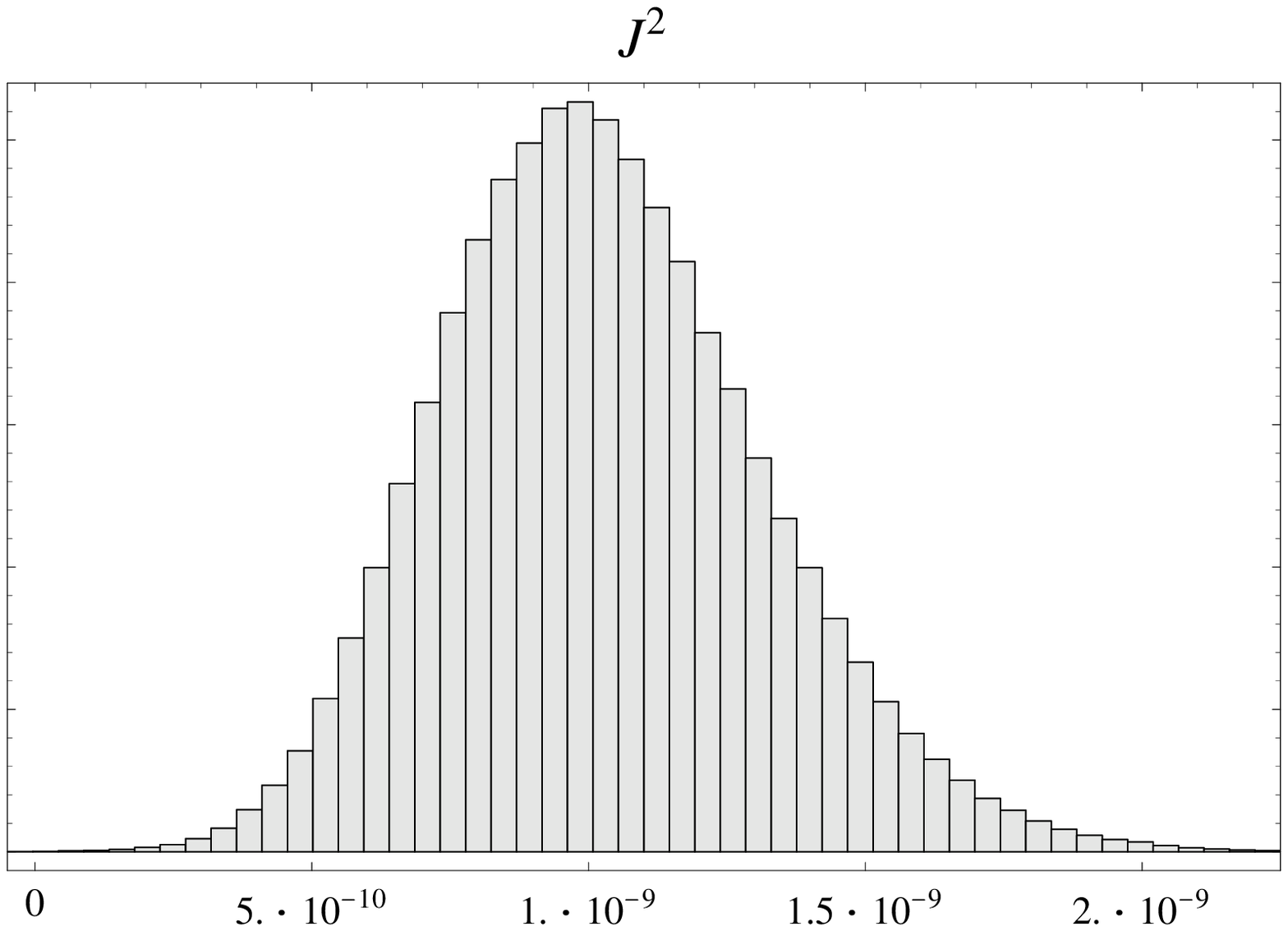,width=0.45\textwidth}}~~\subfigure[$\absV{cd}=0.21955\pm 
0.00002$\label{Fig:03b}]{\epsfig{file=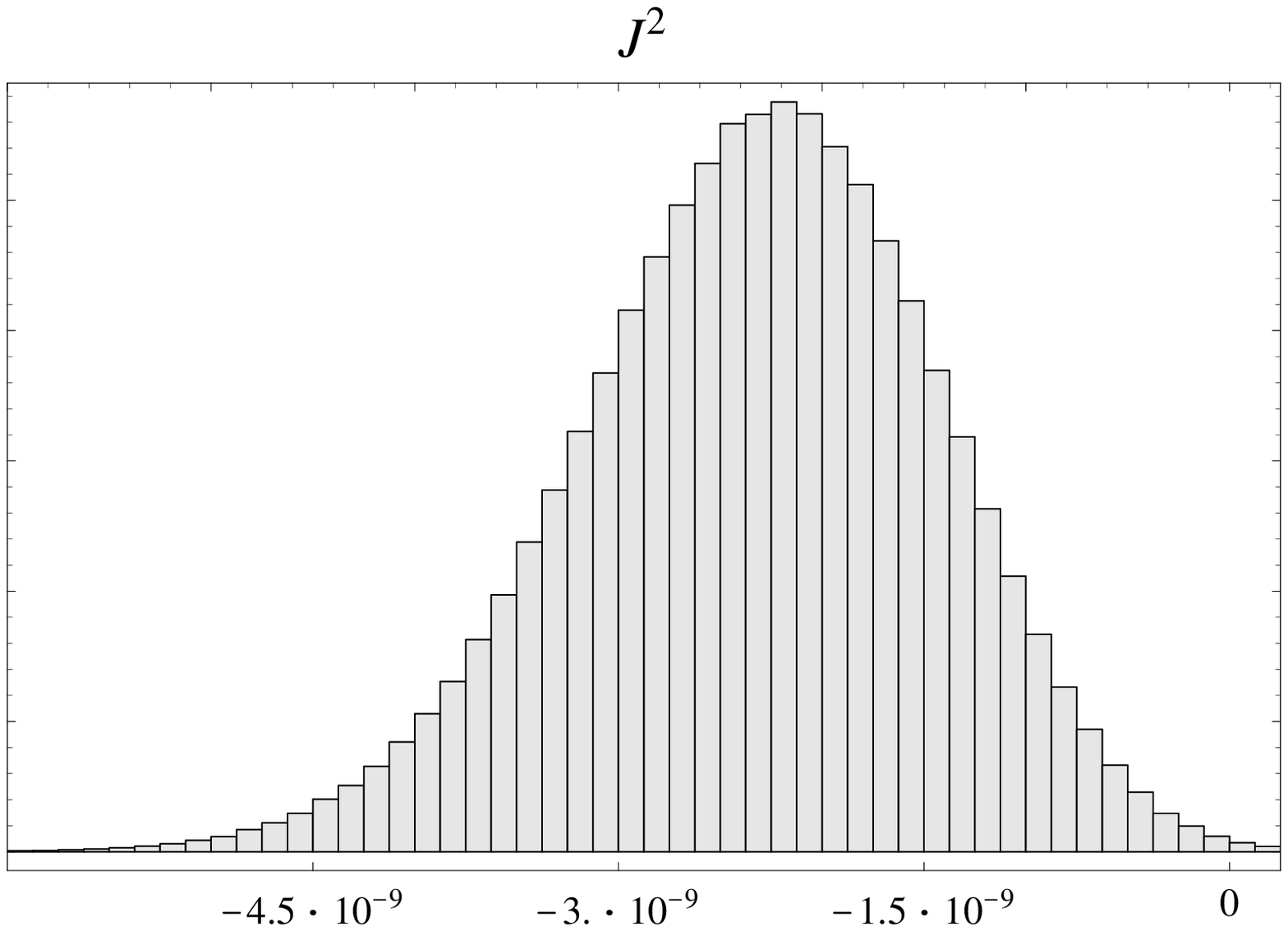,width=0.45\textwidth}}
\end{center}
\caption{$J^{2}$ distributions assuming the unrealistic precision values 
$\absV{us}=0.22000\pm 0.00002$ and $\absV{cd}$.}
\label{Fig:03}
\end{figure}

We wish to emphasize that, contrary to what has been
stated in the literature \cite{Randhawa:2001de},
one cannot interpret the above result as
providing evidence for a complex CKM matrix, using as input data only 
$\absV{us} ,\absV{cb} ,\absV{ub} $ and $\absV{cd}$. This can be
trivially seen from the standard CKM parametrization \cite{Chau:1984fp,PDBook}, 
by noting that from
the input moduli $\absV{us} ,\absV{cb}$ and 
$\absV{ub}$, $s_{12},s_{23}$ and $s_{13}$ can be
obtained. In order to conclude that $\absV{cd}$ has
necessarily a non-vanishing contribution proportional to $\cos \delta _{13}$
(the leading one sensitive to $\delta _{13}$), one would need to know 
$\absV{cd}$ with a relative error of $\leq 10^{-4}$ ! It
is interesting to point out that the leading contribution to 
$\absV{cd}$ comes from $\absV{us}$, therefore
also $\absV{us}$ would have to be known with the same level of
precision, but not $\absV{ub}$ and $\absV{cb}$. 
For completeness we plot in Fig. \ref{Fig:03a} the value of $J^{2}$
extracted from input moduli that are assumed to have the required
unrealistic precision. It is clear that now, essentially 
all points have $J^{2}>0$,
and therefore the corresponding $J=\left( 3.2\pm 0.5\right)\times 10^{-5}$ 
has full meaning. Figure \ref{Fig:03b} shows the $J^2$ distribution 
obtained with the same inputs as in Fig. \ref{Fig:03a}, 
except for the central value of $\absV{cd}$; in this case only $0.16$\% 
of the points have $J^{2}>0$. Notice that both values, 
$\absV{cd}=0.21985\pm 0.00002$ and $\absV{cd}=0.21955\pm 0.00002$, 
are fully compatible with present measurements ($\absV{cd}=0.224\pm 0.012$). 
These two figures clearly show the required precision in $\absV{us}$ 
and $\absV{cd}$ to check 3$\times$3 unitarity and to have an 
irrefutable proof of a complex CKM.

Finally, it should be also emphasized that although the above analysis was made
using the almost collapsed unitarity triangle corresponding to the first two
rows of $V_\mathrm{CKM}$, the same conclusions can be 
obtained if we consider the
``standard triangle'', corresponding to orthogonality of the first and third
columns. One can evaluate any of the angles of the unitarity triangle in
terms of the input moduli, obtaining for example:
\begin{equation}
\sin ^{2}\frac{\gamma }{2}=\frac{\absV{td}^{2}\absV{tb}^{2}-\left( \absV{cd} \absV{cb}
 -\absV{ud} \absV{ub}\right)^{2}}{4\absV{ud} \absV{ub}\absV{cd} \absV{cb}}~,
\label{singamma0}
\end{equation}
Since we are using as input moduli  $\left| V_{us}\right| $, 
$\left| V_{cb}\right| $, $\left| V_{ub}\right| $ and 
$\left| V_{cd}\right| $,
it has to be understood that in Eq. (\ref{singamma0}) one has:
\begin{equation}
\begin{array}{c}
\absV{tb}^{2}=1-\absV{cb}^{2}-\absV{ub}^{2}~, \\ 
\absV{td}^{2}=\absV{us}^{2}-\absV{cd}^{2}+\absV{ub}^{2}~, \\ 
\absV{ud}^{2}=1-\absV{us}^{2}-\absV{ub}^{2}~.%
\end{array}
\label{singamma1}
\end{equation}
The resulting expression can be easily written in terms of the input
moduli as:
\begin{equation}
\cos \gamma =\frac{\absV{cd}^{2}\left( 1-\absV{ub}^{2}\right) -\absV{us}^{2}
\left(1-\absV{cb}^{2}\right) +\absV{ub}^{2}\absV{cb}^{2}}{2\absV{ub}
\absV{cd} \absV{cb} \sqrt{1-\absV{us}^{2}-\absV{ub}^{2}}}~.  \label{cosgamma0}
\end{equation}
It is obvious that a ``legitimate'' 
extraction of $\gamma $ from the input
moduli would require an unrealistic precision in the determination of the
chosen moduli. In particular, the difference $\left|
V_{cd}\right| ^{2}-\left| V_{us}\right| ^{2}$ 
would have to be known at the level
of $\left| V_{us}\right| ^{2}\left| V_{cb}\right| ^{2}$,  
thus requiring 
$\left| V_{cd}\right| $ and $\left| V_{us}\right| $ to be measured with
a $10^{-4}$ relative error. This confirms our previous argument, 
based on the standard parametrization. 
Of course, Eq. (\ref{singamma0})
can be used to determine $\sin ^{2}\gamma /2$ 
if one uses as input moduli $\left| V_{us}\right|$, 
$\left| V_{ub}\right| ,\left| V_{cb}\right| $ and $
\left| V_{td}\right| $. However, $\left| V_{td}\right| $ has the
disadvantage that its extraction from experimental data is
affected by the possible presence of NP. We
will turn to this question in the next section.
                                                                               
Above, we have argued that, although possible in theory, it is not viable in
practice to prove that $V_\mathrm{CKM}$ is non-trivially complex, from the
knowledge of four independent moduli belonging to the first two rows of 
$V_\mathrm{CKM}$. We emphasize that this would be the ideal proof, since the
extraction from experiment of the moduli of the first two rows is
essentially immune to the possible presence of NP.

\section{The impact of $\boldsymbol{\Delta M_{B_{d}}}$ and 
$\boldsymbol{a_{J/\Psi K_{S}}}$ and the NP
parametrization}
\begin{figure}[htb]
\begin{center}
\epsfig{file=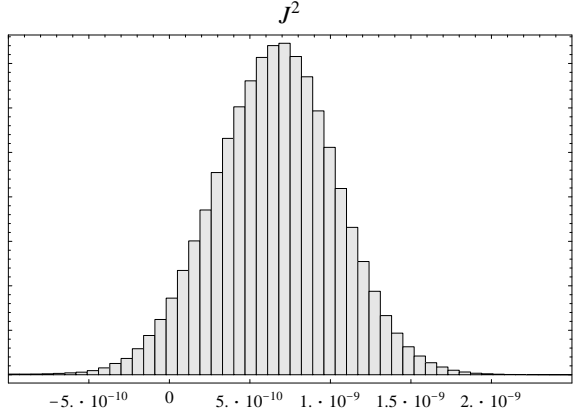,width=0.55\textwidth}
\end{center}
\caption{$J^{2}$ distribution obtained with $\absV{us}$, 
$\absV{cb}$, $\absV{ub}$ and $\absV{td}$ (extracted from $\Delta M_{B_{d}}$).}
\label{Fig:04}
\end{figure}

The experimental measurement of the $B_{d}^{0}$--$\bar{B}_{d}^{0}$
oscillation frequency has reached an average accuracy of almost
$1 \% $ . The theoretical prediction of  $\Delta M_{B_{d}}$ within 
the SM is given by:
\begin{equation}
\Delta M_{B_{d}}= 2 \left| M_{12} \right| = 
\frac{G_F^2M_W^2}{6 {\pi}^2}
\eta_{B_d} m_{B_d} B_{B_d} f^2_{B_d} S_0 (x_t ) \left|
\V{tb} \Vc{td} \right| ^2 \ , 
\label{del}
\end{equation}
where we have followed the standard notation. Equation (\ref{del})
is an approximation valid up to order
$S_0 (x_c, x_t )/ S_0 (x_t ) \sim 10^{-3}$.  From this expression
we can extract $\left|\V{tb} \Vc{td} \right|$. Furthermore, 
unitarity and the experimental values of 
$\left| V_{ub} \right| $ and $\left| V_{cb} \right| $ determine
$\left| V_{tb} \right| $, thus allowing the extraction of 
$\left| V_{td} \right| $.  In Fig. \ref{Fig:04} we plot the new $J^2$
distribution obtained with $\left| V_{td} \right| $, 
thus determined together with $\left| V_{us} \right| $,
$\left| V_{cb} \right| $ and $\left| V_{ub} \right| $. Table \ref{TABLE}
presents all the values used for the input parameters.

There is a striking difference between Fig. \ref{Fig:04} and
Fig. \ref{Fig:01}. As we have emphasized in the previous section,
the majority of points obtained for $J^2$, when one uses as input
$\left| V_{us} \right| $, $\left| V_{cb} \right| $, 
$\left| V_{ub} \right| $, $\left| V_{cd} \right| $,
do not satisfy the unitarity condition that the $uc$
triangle closes (i.e. $J^2 > 0$). On the contrary,
when we use $\left| V_{us} \right| $, $\left| V_{cb} \right| $,
$\left| V_{ub} \right| $, $\left| V_{td} \right| $
to extract $J^2$, the majority of the points satisfies the unitarity
condition that $J^2>0$ and in fact the value of $\left| J \right|$
agrees with its experimental value $\left| J \right| \sim 10^{-5}$
(of course, using exclusively  moduli of $V_\mathrm{CKM}$ one cannot
determine the sign of $J$). This means that, within the SM, one can use
information on $\left| V_{ub}  \right| / \left| V_{cb} \right| $ and
$\Delta M_{B_{d}}$ to conclude that the triangle cannot collapse
to a line, thus implying
a complex CKM matrix. Unfortunately, this is not an irrefutable
proof that CKM is complex, since $\Delta M_{B_{d}}$ can receive
significant contributions from NP, 
as we have emphasized in section \ref{sec:02}.
The same is true for the CP asymmetry $a_{J/\Psi K_{S}}$ in the 
$B_d^0,\bar B_d^0\to J/\Psi K_S $ decays.
Let us introduce the following parametrization of the NP contribution
to $B_{d}^{0}$--$\bar{B}_{d}^{0}$ mixing:
\begin{equation}
M_{12}=r_d^2 e^{-i2\phi_d}[M_{12}]_\mathrm{SM}
\label{m12}
\end{equation}
where $[M_{12}]_\mathrm{SM}$ is the SM box diagram contribution. 
The NP contribution to $M_{12}$ implies that 
$a_{J/\Psi K_{S}}$ no longer measures $2 \beta$, but one has instead:
\begin{equation}
a_{J/\Psi K_{S}}=\sin 2(\beta-\phi_d)=\sin 2\overline\beta~.
\label{ajp}
\end{equation}
Although the expression for $[M_{12}]_\mathrm{SM}$ 
is the one given by the SM,
its actual numerical value may differ from the SM prediction, since
models beyond the SM allow in general for a different range
of the CKM matrix elements.

It is clear that allowing for the presence of NP in $M_{12}$,
parametrized as in Eq. (\ref{m12}), one can fit the data on 
$a_{J/\Psi K_{S}}$, $\Delta M_{B_{d}}$, even with a real CKM
matrix, by putting:
\begin{equation}
\beta = 0; \qquad \sin (2\phi_d)= -a_{J/\Psi K_{S}}
\label{s2f}
\end{equation}
and then adjusting $r_d$ to fit $\Delta M_{B_{d}}$. Obviously,
two solutions are obtained for $r_d$, corresponding to the
two ways the unitarity triangle can collapse (i.e. $\gamma = 0$ or
$\gamma = \pi$). It must be stressed that to write 
Eq. (\ref{ajp}) one has to assume that the potential NP 
contamination to the penguin diagram contributing to the decay 
amplitude is negligible. This process is dominated by a tree and 
a penguin with the same phase; therefore, in this particular case, 
it is natural to assume no NP in the decay amplitude. 
For very small deviations from these assumptions see \cite{Atwood:2003tg}.

\begin{figure}[htb]
\begin{center}
\subfigure[Apex $-\frac{V_{ud}V_{ub}^\ast}{V_{cd}V_{cb}^\ast}$ 
of the unitarity triangle $db$.\label{Fig:05a}]
{\epsfig{file=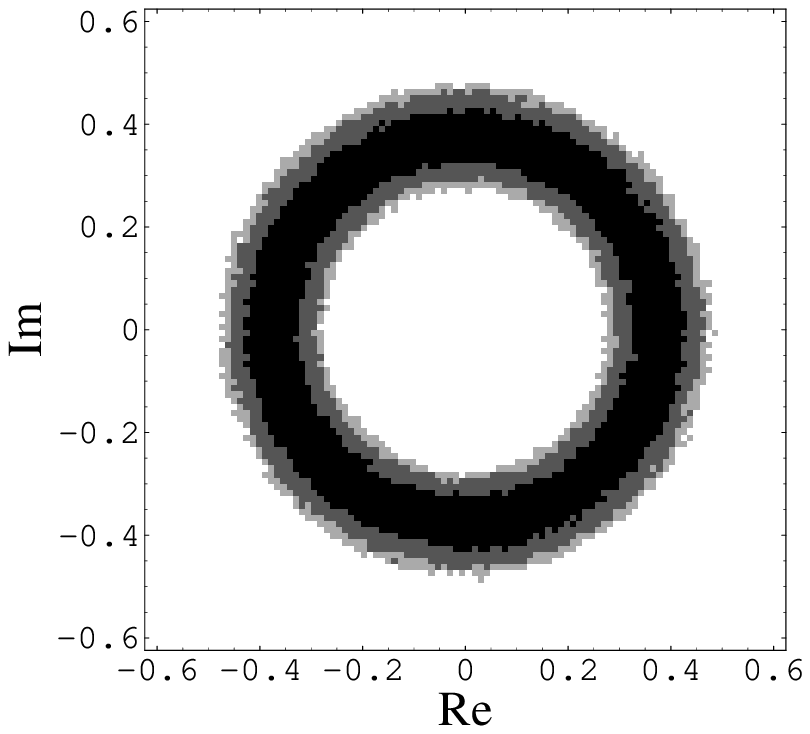,width=0.45\textwidth}}~~\subfigure[$(r_d^2,2\phi_d)$ 
joint distribution.\label{Fig:05b}]{\epsfig{file=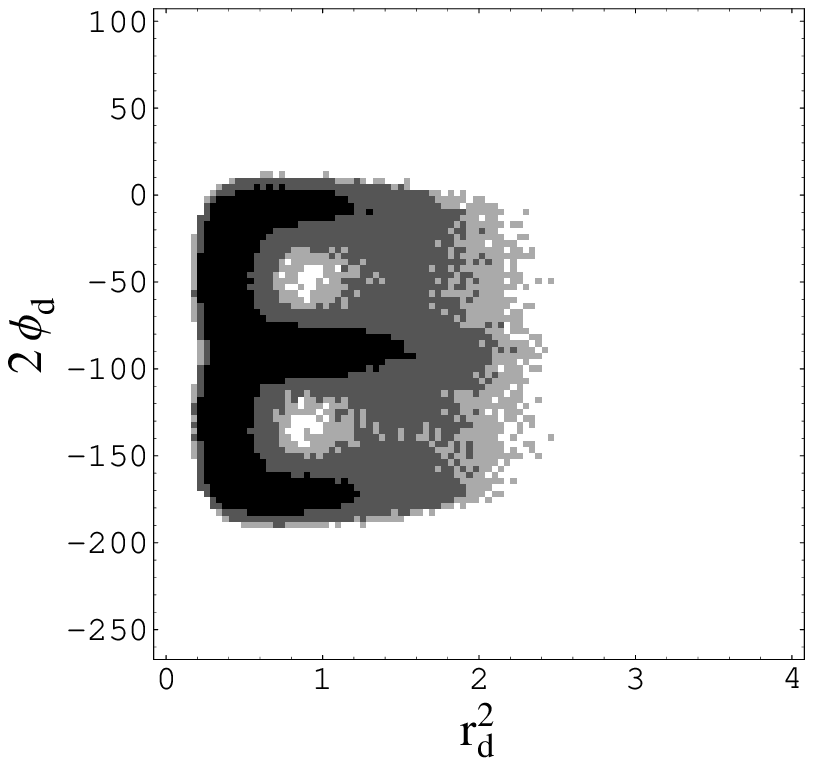,width=0.45\textwidth}}
\end{center}
\caption{68\%(black), 90\%(dark grey) and 95\%(grey) probability regions.}
\label{Fig:05}
\end{figure}

Independently of the question of whether the CKM matrix is complex 
or not, it is interesting to 
investigate what the allowed values for
$r_d$, $\phi_d$ are, taking into account the present data, summarized in
Table \ref{TABLE}. We calculate, using again the Monte Carlo method, 
probability density distributions for CKM parameters, 
$r_d^2$ and $2\phi_d$ using as constraints Gaussian distributions 
for moduli of the first two rows of the CKM matrix, $\Delta M_{B_{d}}$ 
and $a_{J/\Psi K_{S}}$. In Fig. \ref{Fig:05a} we plot 68\% (black), 90\% 
(dark grey) and 95\% (grey) probability regions of the probability 
density function (PDF) of the apex $- \V{ud}\Vc{ub} / \V{cd}\Vc{cb}$ 
of the $db$ unitarity triangle. In Fig. \ref{Fig:05b} we represent 
joint PDF regions in the plane $(r^2_d$, $2\phi_d)$. 
Because $\gamma$ gives the apex of the triangle, it is clear 
from Fig. \ref{Fig:05a} that there is essentially no restriction 
on $\gamma$. On the contrary, because the moduli of the first two 
rows put an upper bound on $|\beta|$ and upper and lower bounds 
on $R_t=\abs{\V{td}\Vc{tb}}/\abs{\V{cd}\Vc{cb}}$, 
we can see in Fig. \ref{Fig:05b} 
significant constraints on $2\phi_d$ and $r_d^2$. 
Although the experimental value of the semileptonic asymmetry 
$A_\mathrm{SL}$ (the asymmetry in the number of equal 
sign lepton pairs arising from the semileptonic decay of
$B_{d}^{0}$--$\bar{B}_{d}^{0}$ pairs)
has some impact on these figures, we have not included 
it in order to show in a clear way the impact of the actual 
measurements. We will come back to this point later on.

At this stage, the following comment is in order. We are
assuming a class of theories beyond the SM that give NP
contributions to  $\Delta M_{B_{d}}$, but keep $V_\mathrm{CKM}$
unitary. The most important example of this class of theories
are the supersymmetric extensions of the SM. Since in this
framework $V_\mathrm{CKM}$ is still unitary, once  
$\left| V_{us} \right| $, $\left| V_{cb} \right| $, 
$\left| V_{ub}  \right| / \left| V_{cb} \right| $ are fixed 
by data, there is only $\beta$ as a free independent CKM
parameter. It is then clear that with only two experimental inputs,
namely  $\Delta M_{B_{d}}$, $a_{J/\Psi K_{S}}$,
one cannot fix the three parameters $r_d$, $\phi_d$, $\beta$.
One may be tempted to include information on
$\Delta M_{B_{s}}$. However, in our scenario, this would require the
introduction of new parameters ($r_s$, $\phi_s$) giving NP 
contributing to $\Delta M_{B_{s}}$.

In the next section, we will emphasize the importance of $\gamma$
in obtaining an irrefutable proof that $V_\mathrm{CKM}$
is complex, even allowing for the presence of NP.

\section{The importance of a direct measurement of the phase 
$\boldsymbol{\gamma}$}

In our framework, we can distinguish two conceptually different ways of
measuring $\gamma$ in $B_{d}^{0}$ non-leptonic decays: those that come from the
interference of two \emph{tree-level} decays 
such as $b\rightarrow s\bar{u}c$
with $b\rightarrow s\bar{c}u$, or $b\rightarrow d\bar{u}c$ with $%
b\rightarrow d\bar{c}u$, and those other processes where the presence of 
\emph{both tree and penguin} diagrams is allowed, such as $b\rightarrow d%
\bar{u}u$ interfering with $b\rightarrow d$. The first group includes
the usual methods that are called, in the literature,  
methods to measure $%
\gamma $  or $2\overline{\beta }+\gamma $ \cite{Gronau:1990ra,Giri:2003ty}.
In general they involve the
transitions $B\rightarrow DK$ or $B\rightarrow D\pi $ as the 
flavour-representative examples of the considered decays. 
The second group includes
the usual methods to measure $\alpha $ \cite{Gronau:1990ka,Snyder:1993mx},
or more generally 
$\overline{\alpha }=\pi -\overline{\beta }-\gamma $ 
if one allows for the presence of NP in $B_{d}^{0}$--$\bar{B}_{d}^{0}$ mixing. 
By now it is well known that the
so-called methods to measure $\overline{\alpha }$ in fact ought to be called
methods\footnote{
Note that by definition $\alpha +\beta +\gamma =\pi $ and also 
$\overline{\alpha }+
\overline{\beta }+\gamma =\pi $. This relation does not test $3\times 3$
unitarity, as it cannot be violated since it is a definition. Recall that
with $3$ complex numbers (the sides of the unitarity triangle -- even if they
do not close --) we can define only two independent relative angles: the third
one is always a combination of the other two.} 
to measure $\overline{\beta }+\gamma $ \cite{Silva:2004gz,Barenboim:1997qx}.
Of course, the
representative example of this method as far as flavour content is concerned
is $B\rightarrow \pi \pi $.

It has to be stressed that there are other available methods to extract $%
\gamma $, i.e. using  $b\rightarrow s$ transitions. However, in the search
for NP, great attention has to be payed to the assumptions made in order to
keep the errors under control.
In this respect, $SU(2)$ is a very good symmetry
of the strong interactions, but the same does not hold for $SU(3)$.  
Furthermore, there are
no first principle calculations of $SU(3)$ breaking effects in
non-leptonic $B$-decays. Therefore, 
although analyses based on $SU(3)$, $U$-spin,
can be quite useful in some instances, we will restrict ourselves to 
$SU(2)$-based analyses, as far as strong interactions are concerned.

Concerning the two previous methods, their fundamental differences lie in
the presence or not of penguin diagrams. The method to extract $\gamma $ or $%
2\overline{\beta }+\gamma $ relies on pure tree-level processes, therefore
no additional assumptions are needed to get a value of $\gamma \neq 0$ so
as to have an irrefutable proof for a complex $V_\mathrm{CKM}$. Of course
it has to be understood that $2\overline{\beta }$ has been extracted from $%
{B}_{d}^{0},\bar{B}_{d}^{0}\rightarrow J/\Psi K_{S},
J/\Psi {K} ^{0\ast },J/\Psi \bar{K} ^{0\ast }$ in order to use the 
method of the $2%
\overline{\beta }+\gamma $ extraction. In the $\overline{\alpha }$ method,
the presence of penguins in general calls for the additional assumption of
no NP competing with these penguins in the decay amplitudes. Therefore we
will present two separate analyses with the actual relevant data.

\subsection{$\boldsymbol{\gamma}$ from pure weak tree-level decays}

The enormous effort developed at the
B-factories Belle and BaBar has resulted in
the first measurements of $\gamma$ -- although by now 
still poor -- in tree-level
decays $B^{\pm }\rightarrow DK^{\pm }$, $B^{\pm }\rightarrow D^{\ast }K^{\pm
}\rightarrow \left( D\pi ^{0}\right) K^{\pm }$,  
where the two paths to $D^{0}$ or $\bar{D}^{0}$ interfere 
in the common decay channel $\bar{D}^{0},{D}^{0}\rightarrow K_{S}\pi ^{+}
\pi ^{-}$. From a Dalitz-plot 
analysis \cite{Abe:2004gu}, Belle has presented 
$\gamma =68^\circ\pm\begin{smallmatrix}14^\circ\\ 15^\circ\end{smallmatrix}
\pm 13^\circ\pm 11^\circ$ and BaBar \cite{Aubert:2004kv} $\gamma 
=70^\circ\pm 26^\circ\pm 10^\circ\pm 10^\circ$, together with 
the solutions obtained by changing $\gamma \rightarrow\gamma \pm \pi $. 
The statistical significance of CP violation in the Belle
measurement is of $98\%$ \cite{Abe:2004gu}; by now therefore this can be
taken as an indicative figure of the statistical significance of having a
complex CKM.

The method used by Belle and BaBar suffers from model-dependent assumptions
in the Dalitz-plot analysis; these can be eliminated with more statistics
following the suggestion of Ref. \cite{Giri:2003ty}. Also, the presence of
NP in the $D^{0}$--$\bar{D}^{0}$ mixing could modify this analysis, but if
necessary, in case of better precision, 
this small correction \cite{Giri:2003ty}
could be included in the way suggested in Refs.
\cite{Amorim:1998pi}.

\begin{figure}[htb]
\begin{center}
\subfigure[Apex $-\frac{V_{ud}V_{ub}^\ast}{V_{cd}V_{cb}^\ast}$ of the 
unitarity triangle $db$.
\label{Fig:06a}]{\epsfig{file=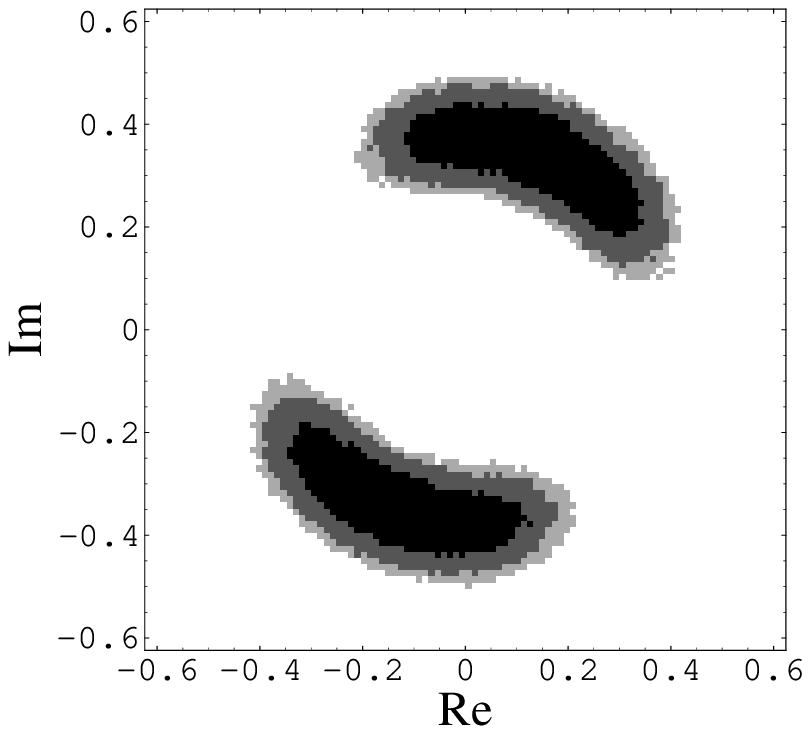,width=0.45\textwidth}}~~\subfigure[$(r_d^2,2\phi_d)$ joint distribution.\label{Fig:06b}]{\epsfig{file=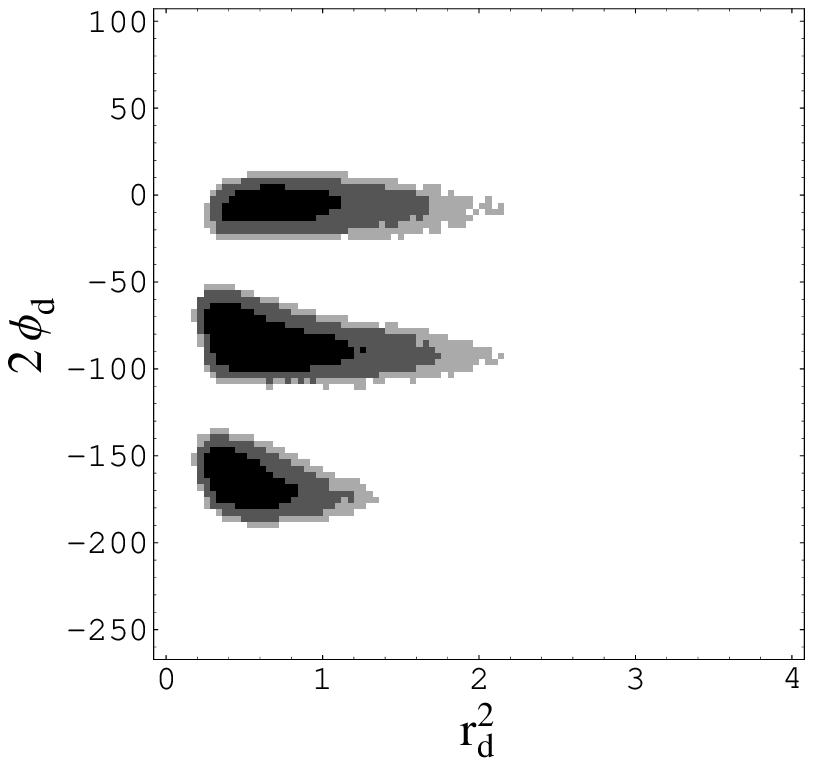,width=0.45\textwidth}}
\end{center}
\caption{68\% (black), 90\% (dark grey) and 95\% (grey) probability 
regions, including the constraint on $\gamma$.}
\label{Fig:06}
\end{figure}

We average conservatively both measurements to the value 
$\gamma =69^\circ\pm 21^\circ$ ($-111^\circ\pm 21^\circ$), 
which we take as a quantitative measurement of a complex CKM matrix 
\emph{independent of the presence of NP at the one-loop weak level}. 
To analyse the implications for the presence of NP, we add this new 
constraint to the previous analysis presented in Figs. 
\ref{Fig:05a} and \ref{Fig:05b}. In Fig. \ref{Fig:06a} we represent 
the analogue of Fig. \ref{Fig:05a}.
From Table \ref{sols:01}, where we only present 
the central values, we clearly see that with two $\gamma$ values and two signs for $\cos(2\overline{\beta }) $ we generate four solutions.

It has to be stressed that although $\overline{\beta }$ presents a fourfold
ambiguity, $2\overline{\beta }$ -- necessary to extract $2\phi _{d}$ --
presents a twofold one. With a measurement of $\gamma $ (another twofold
ambiguity) we have $\beta $ and $R_{t}$ fixed up to a twofold ambiguity, and
therefore  in total we have four different solutions 
for $r_{d}^{2}$, $2\phi _{d}$ and the CKM parameters.

These solutions are represented in the $(r_{d}^{2},2\phi _{d})$ plane as the
joint PDF in Fig. \ref{Fig:06b}. We have the ``SM solution'' corresponding to 
$2\phi_d \sim 0^\circ$, $r_d\sim 1$, two NP solutions overlapping at 
$2\phi _{d}\sim -85^\circ$ and 
another NP solution near $2\phi _{d}\sim -169^\circ$. 
Had we used the constraint on the sign of $\cos (2\overline{\beta })$,
from $B_{d}^{0}\rightarrow J/\Psi K^{\ast }$, we would have eliminated two
NP solutions, but the situation is not conclusive (see \cite{BabarSignCos2Beta:2004}). 
The fact that the
central value of $\gamma $ is a little bit high with respect to the 
value obtainable in a SM fit, implies a large value for $R_{t}$ 
for the ``SM solution'', and therefore a low central value of $r_d$, i.e. $r_{d}^{2}<1$.

Information on $2\overline{\beta }+\gamma $ is also available from partially
reconstructed $B^{0}\rightarrow D^{\ast \pm }\pi ^{\mp }$ decays, $\left\vert
\sin \left( 2\overline{\beta }+\gamma \right) \right\vert >0.58$ (90\% C.L.)
\cite{Aubert:2004pt}. But at the moment this result relies on
theoretical inputs based on $SU(3)$, and we will therefore not use this
constraint.
Coming back to our irrefutable proof of a complex CKM, with our PDF for $\gamma$ 
we find that a real CKM matrix is excluded at a 99.84\% C.L.\footnote{With the standard construction used by Babar and Belle this C.L. is obtained as the integrated probability over the values of $\gamma$ having more probability than the most probable CP-conserving value of $\gamma$ \cite{fernando}.}. In much more intuitive terms, we find that the probability corresponding to $\gamma \in [10^\circ;170^\circ]\cup [-170^\circ;-10^\circ]$ is 99.7\%.

\begin{center}
\begin{table}
\begin{tabular}{|c|c|c|c|c|c|c|c|}
\hline
 $\gamma$  {\tiny (input)} & $\beta$ & $2\phi_d$ & $r_d^2$ & $R_t$ 
& Sign[cos($2\overline\beta$)]&$2\overline\beta$ {\tiny (input)} 
& $\overline \alpha$\\ \hline\hline
 $69^\circ$ &$23.2^\circ$& $-0.8^\circ$& $0.80$& $0.93$& + & $47.2^\circ$& $87.4^\circ$\\ \hline
 $69^\circ$ &$23.2^\circ$& $-86.4^\circ$ & $0.80$& $0.93$& - & $132.8^\circ$ & $44.6^\circ$ \\ \hline
 $-111^\circ$&$-17.9^\circ$ & $-83.0^\circ$&$0.48$ &$1.20$ & + & $47.2^\circ$ &$-92.6^\circ$ \\ \hline
 $-111^\circ$&$-17.9^\circ$ & $-168.6^\circ$&$0.48$ &$1.20$ & - & $132.8^\circ$&$-135.4^\circ$ \\ \hline
\end{tabular}
\caption{``Central'' values for the solutions (see Fig. \ref{Fig:06}).}\label{sols:01}
\end{table}
\end{center}

\subsection{$\boldsymbol{\gamma}$ from $\boldsymbol{ \overline \alpha}$ methods}

Using these $\overline{\alpha }$ methods to get an irrefutable proof 
of a complex $
V_\mathrm{CKM}$ is in principle much more subtle. They are all 
based on the transition 
$b\rightarrow du\bar{u}$, so that the so-called 
penguin pollution (weak loops) can also become NP pollution. 
The most prominent channels are $B\rightarrow \pi \pi $, $B\rightarrow \rho \pi $ and $B\rightarrow \rho \rho$. 
%
In the appendix we have shown that the extraction of $\overline\alpha=\pi-\overline\beta-\gamma=\alpha+\phi_d$
from these channels is valid even in the presence of any NP in the $\Delta I=1/2$ piece of the weak hamiltonian. 
As the $\Delta I=3/2$ piece is dominated by tree diagrams, this extraction of $\overline\alpha$ is in accordance with our initial assumptions: ``no NP in weak processes where the SM contributes at tree level'' means in this case no NP in the $\Delta I=3/2$ piece. We conclude that the extraction of $\overline\alpha$ from these channels is completely independent of the presence of NP in the usual penguins. Only NP in electroweak penguins (EWP) could spoil this conclusion, but this is very unlikely due to the small effect of the SM EWP, of order $1.5^\circ$ (see reference \cite{0502139Gronau}), that can be eventually taken into account.

BaBar has presented a time-dependent analysis of the $\rho ^{+}\rho ^{-}$
channel \cite{Aubert:2004zr}, that once supplemented with the $\rho ^{+}\rho
^{0}$ and $\rho ^{0}\rho ^{0}$ branching ratios 
\cite{Aubert:2003wr,Zhang:2003up}  and the measurement 
of the final polarization \cite{Aubert:2004zr}, can be 
translated \cite{Bevan:2004ht} into the
measured value 
$\overline{\alpha }=96^\circ\pm 10^\circ\pm 5^\circ\pm 11^\circ$ 
where the last error comes from the usual $SU(2)$ 
isospin bounds\footnote{In our notation $\overline{\alpha }_\mathrm{eff}$ 
is the usual $\alpha _\mathrm{eff}$ of Refs. 
\cite{Gronau:1990ka,Charles:2004jd,Silva:2004gz},
but where we have introduced $\overline{\beta }$ 
instead of $\beta $ as the phase in the 
$B_{d}^{0}$--$\bar{B}_{d}^{0}$ mixing.} 
$\left\vert \overline{\alpha }_\mathrm{eff}-\overline{\alpha }\right\vert 
\leq 11^\circ$. Because the measurement is sensitive to 
$\sin ( 2\overline{\alpha }_\mathrm{eff}) $, 
$\overline{\alpha }=\overline{\alpha }_\mathrm{eff}\pm 11^\circ$ 
presents a fourfold ambiguity $\left( \overline{\alpha },\overline{\alpha }%
+\pi ,\frac{\pi }{2}-\overline{\alpha },-\overline{\alpha }-\frac{\pi }{2}%
\right) $. In the $\rho \pi $ channel the 
pentagon isospin analysis -- from quasi-two-body
decays -- needs more statistics and/or additional assumptions. A
time-dependent Dalitz-plot analysis in the channel 
$B\rightarrow \pi ^{+}\pi
^{-}\pi ^{0}$ has been presented by BaBar 
\cite{Aubert:2003wr,Aubert:2004iu}, with the result 
$\overline{\alpha }=113^\circ\pm
\begin{smallmatrix}27^\circ\\ 17^\circ\end{smallmatrix}\pm 6^\circ$.

Since this analysis is sensitive to both $\sin \left( 2\overline{\alpha }%
_\mathrm{eff}\right) $ and 
$\cos \left( 2\overline{\alpha }_\mathrm{eff}\right) $, the
resulting ambiguity is just a twofold one $\left( \overline{\alpha },%
\overline{\alpha }+\pi \right) $. It is remarkable that these two solutions
are in good agreement with two of the solutions coming from the $\rho \rho $
channel. This important property will be used to eliminate two of the four
solutions coming from the $\rho \rho $ channel, in order to see 
the future trend of this analysis in a much
clearer way.

The situation in the $\pi \pi $ channel does not yet allow 
a full isospin analysis
and the isospin bounds are quite poor. Furthermore  BaBar and 
Belle measurements are
still in some conflict, so that we will not use these results.

As before, we average the data from $\rho \rho $ and $\rho \pi $ but only 
keep the two solutions consistent with the $\rho \pi $ channel data.
Our averaged values are 
$\overline{\alpha }=100^\circ\pm 16^\circ$, ($-80^\circ\pm 16^\circ$).

\begin{figure}[htb]
\begin{center}
\subfigure[Apex $-\frac{V_{ud}V_{ub}^\ast}{V_{cd}V_{cb}^\ast}$ of 
the unitarity triangle $db$.
\label{Fig:07a}]{\epsfig{file=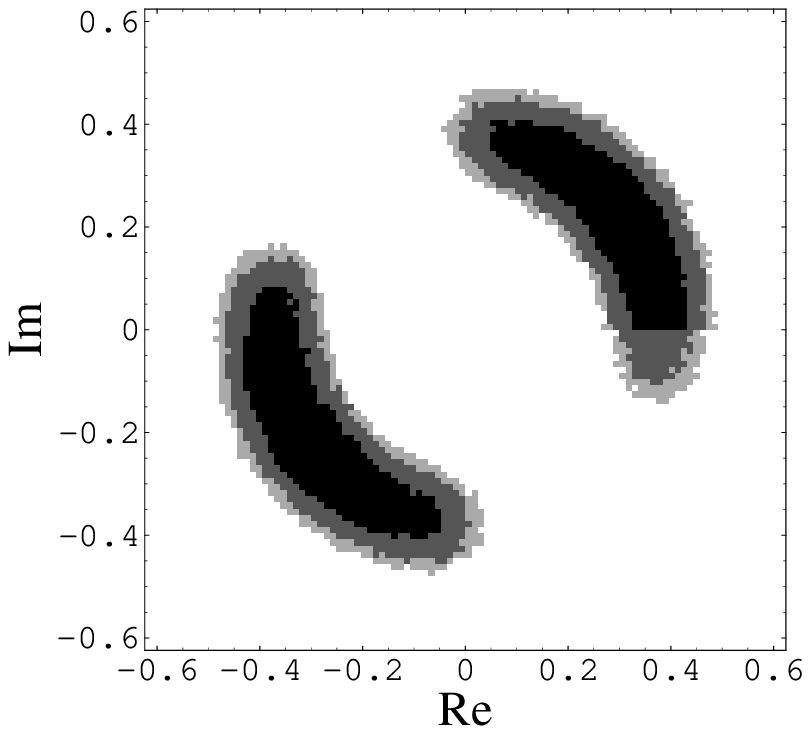,width=0.45\textwidth}}~~\subfigure[$(r_d^2,2\phi_d)$ joint distribution.\label{Fig:07b}]{\epsfig{file=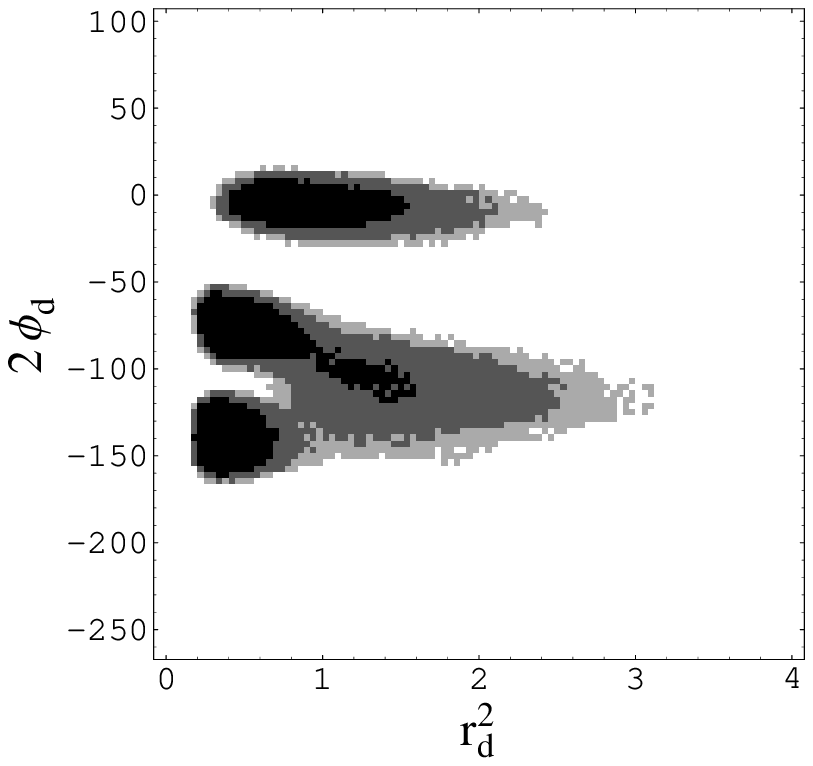,width=0.45\textwidth}}
\end{center}
\caption{68\% (black), 90\% (dark grey) and 95\% (grey) probability regions, 
including the constraint on $\gamma$ through $\overline \alpha$.}
\label{Fig:07}
\end{figure}

The analogue of Fig. \ref{Fig:06a} for the apex of the unitarity triangle,
with the information on $\gamma $ replaced by 
that on $\overline{\alpha }$,
is represented in Fig. \ref{Fig:07a}. 

In this case the data are not conclusive enough to eliminate 
a real $V_\mathrm{CKM}$; a real CKM is excluded at a 75\% C.L.. In the second approach, the probability of $\gamma\in[10^\circ;170^\circ]\cup[-170^\circ;-10^\circ]$ is 85\%. In Table \ref{sols:02},
where we include the central values, we have four solutions from the
two $\overline{\alpha }$ values and the two signs of 
$\cos ( 2\overline{\beta })$. Again, the fourfold $\overline{\beta }$ 
ambiguity is effectively reduced to
two because to reconstruct $\gamma $, the quantity that matters 
is $(\overline{\alpha }+\overline{\beta }) =\pi -\gamma $. 
These solutions are represented in Fig. \ref{Fig:07b}. As before, we have
the SM solution and the other three NP solutions that tend to 
overlap here more than in the previous case.

\begin{center}
\begin{table}
\begin{tabular}{|c|c|c|c|c|c|c|c|}
\hline
 $\gamma$ & $\beta$ & $2\phi_d$ & $r_d^2$ & 
$R_t$ & Sign[cos($2\overline\beta$)]&$2\overline\beta$ 
{\tiny (Input)} & $\overline \alpha$ {\tiny (Input)}\\ \hline\hline
$56.4^\circ$ & $22.8^\circ$  & $-1.7^\circ$ & $0.96$ & $0.85$ 
& + &$47.2^\circ$ & $100^\circ$\\ \hline
$13.6^\circ$ & $8.5^\circ$ &$-115.7^\circ$  & $1.78$ & $0.62$ 
& - &$132.8^\circ$ & $100^\circ$\\ \hline
$-123.6^\circ$ & $-15.1^\circ$ & $-77.4^\circ$ & $0.44$ & $1.26$ & + 
&$47.2^\circ$ &$-80^\circ$ \\ \hline
$-166.4^\circ$ & $-3.8^\circ$ & $-140.5^\circ$ & $0.36$ & $1.39$ 
& - &$132.8^\circ$ &$-80^\circ$ \\ \hline
\end{tabular}
\caption{``Central'' values for the solutions (see Fig. \ref{Fig:07}).}\label{sols:02}
\end{table}
\end{center}
Comparing the two tables and taking errors into account, it is easy to
recognize that the two solutions with positive sign of 
$\cos (2\overline{\beta })$ 
can be consistent, but the negative ones are somewhat inconsistent. 
The reason can be seen by looking at the $\overline{\alpha }$
generated from the $\gamma $ measurement. Because $\overline{\alpha }=\pi -
\overline{\beta }-\gamma $, the two $\overline{\alpha }$ generated with the
same $\overline{\beta }$ (the same sign of $\cos (2\overline{\beta })$) differ
by $\pi $, so if one of the $\gamma$ measurements is 
consistent with one of the $\overline{\alpha }$ measurements, 
so is the other, because both measurements
of $\overline{\alpha }$ and $\gamma $ differ by $\pi $. On the contrary, for
a fixed $\gamma $ measurement, the  two $\overline{\alpha }$ generated with
the different $\overline{\beta }$ (different sign of $\cos (2\overline{\beta })
$) differ by $42.8^\circ$, so if one solution 
is compatible with the $\overline{\alpha }$
measurement, the other is not. In our scenario, therefore, a way 
of measuring the sign of $\cos (2\overline{\beta })$ relies 
on the measurements of $\sin (2\overline{\beta })$,
 $\sin (2\overline{\alpha })$, $\cos (2\overline{\alpha })$ 
and $\tan \gamma $. 

\begin{figure}[htb]
\begin{center}
\subfigure[Apex $-\frac{V_{ud}V_{ub}^\ast}{V_{cd}V_{cb}^\ast}$ 
of the unitarity triangle $db$.
\label{Fig:08a}]{\epsfig{file=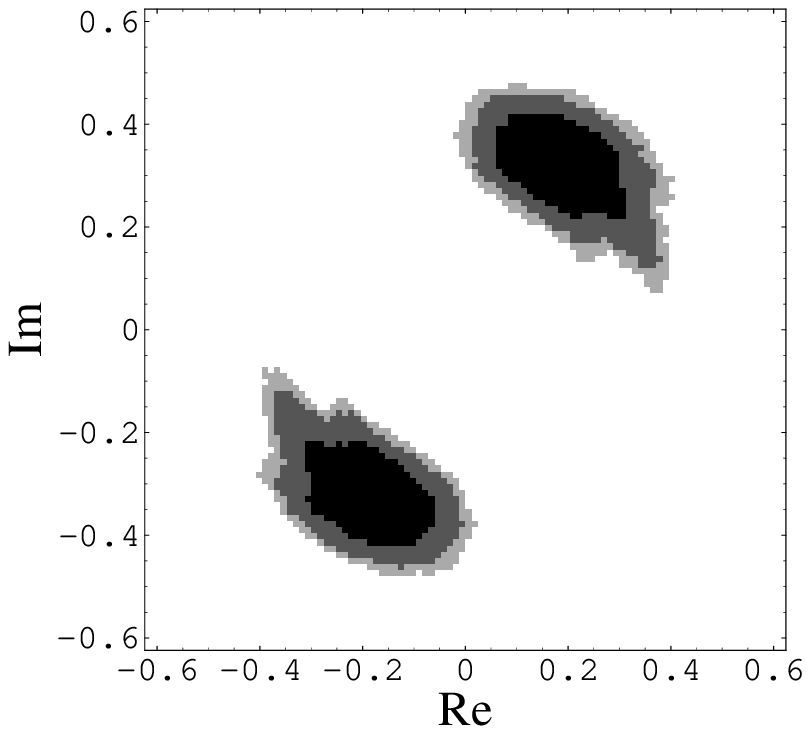,width=0.45\textwidth}}~~\subfigure[$(r_d^2,2\phi_d)$ 
joint distribution.\label{Fig:08b}]{\epsfig{file=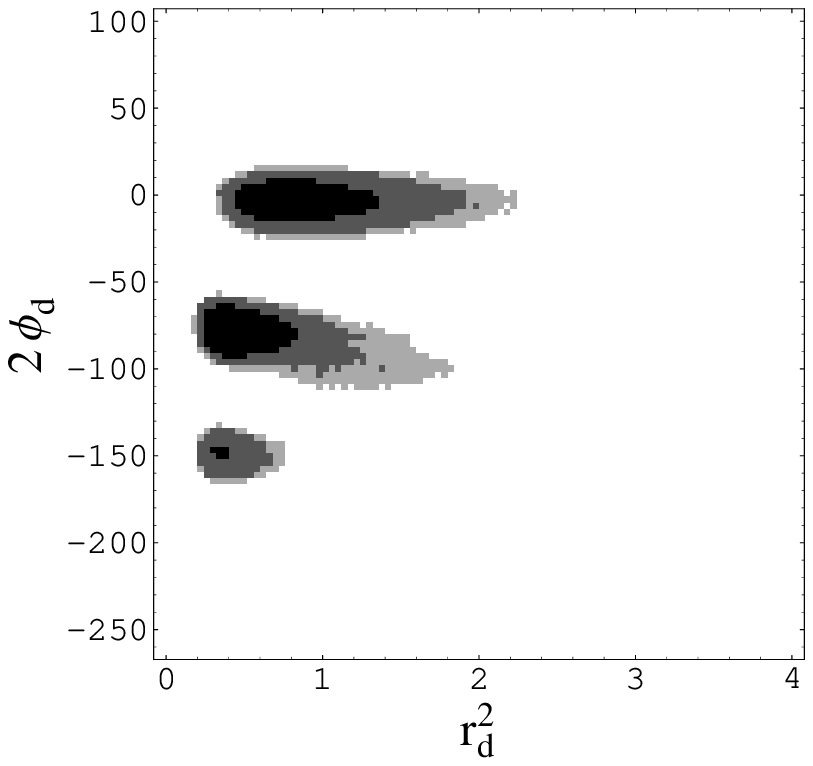,width=0.45\textwidth}}
\end{center}
\caption{68\% (black), 90\% (dark grey) and 95\% (grey) probability 
regions, including the constraints on $\gamma$ derived from both direct 
$\gamma$ measurements as well as from $\overline \alpha$.}
\label{Fig:08}
\end{figure}

In Fig. \ref{Fig:08}  we repeat the previous analyses including 
both constraints from $\gamma $ and $\overline{\alpha }$. 
As can be seen the solutions with
negative $\cos (2\overline{\beta })$ have almost disappeared, 
in fact just $11.1\%$ of the points  have negative 
$\cos (2\overline{\beta })$, contrary to the
previous plots, where they were $50\% $. Combining $\gamma$ and $\overline\alpha$ measurements, a real CKM matrix is excluded at a 99.92\% C.L. and the probability of $\gamma\in[10^\circ;170^\circ]\cup[-170^\circ;-10^\circ]$ is 99.7\%.

As we have mentioned, previous work on these analyses of New Physics has been presented by Z. Ligeti in \cite{Ligeti:2004ak}. For similar experimental inputs and theoretical assumptions the results agree.

\subsection{Further comments}

The semileptonic asymmetry is 
$A_\mathrm{SL}=\text{Im}\left(\Gamma_{12}/M_{12}\right)$, 
where in our scenario $M_{12}$ is polluted by NP, 
$M_{12}=\frac{1}{2}\Delta M_{B_d}e^{i2\overline \beta}$, 
but -- to the present level of precision -- 
$\Gamma_{12}\propto(a+be^{-i\gamma}+ce^{-i2\gamma})$ 
is not; $a,b,c$ depend on QCD parameters and 
moduli of the first two rows of the CKM matrix 
\cite{Beneke:2003az,Branco:1992uy,Charles:2004jd}.
Therefore, by including the measurements of 
$\Delta M_{B_d}$, $a_{J/\Psi K_{S}}$ and a determination of the 
sign of $\cos(2\overline\beta)$, $A_\mathrm{SL}$ could be used in the 
future to obtain information on $\gamma$. The impact of the actual 
value $A_\mathrm{SL}=(-3\pm 7)\times 10^{-3}$ \cite{Group:2004cx} in 
Figs. \ref{Fig:06}, \ref{Fig:07} and \ref{Fig:08} is very mild, 
but a future determination of $A_\mathrm{SL}$ at the level of the SM 
precision ($A_\mathrm{SL}^\mathrm{SM}=(-5.0\pm 1.1)\times 10^{-4}$
\cite{Beneke:2003az} 
) would be sufficient by itself to completely eliminate 
some $\gamma$ solutions.

In some parts of our analyses we have used 3$\times$3 unitarity. 
It is important to clarify that a direct measurement of 
$\gamma$ from tree-level decays (or in the future from $A_\mathrm{SL}$) 
provides an irrefutable 
proof of a complex CKM matrix, independent of violations of 3$\times$3 
unitarity; this includes both $\gamma$ and $\overline\alpha$ methods. This is not the case of our analyses of NP, where we have 
extensively used 3$\times$3 unitarity, for example in the 
extraction of $\beta$ and $R_t$ from moduli of the first two 
rows and $\gamma$ so as to obtain $2\phi_d$ and $r_d^2$.

\section{Summary and Conclusions}
We have carefully examined the present experimental evidence
in favour of a complex CKM matrix, even allowing for NP contributions
to $\epsilon _{K}$, $\Delta M_{B_{d}}$, $\Delta M_{B_{s}}$, $a_{J/\Psi K_S}$, $B\to\rho\rho$ and $B\to\rho\pi$.
First, we showed that, contrary to some statements in the literature
\cite{Randhawa:2001de},
it is not feasible in practice to derive a non-vanishing
value of $|J|$ using only the present information on moduli of the 
first two rows of the CKM matrix. 
We showed this by pointing out that a legitimate determination
of $|J|$ from these moduli, although possible in theory, would
in practice require a completely unrealistic high-precision 
determination of $\absV{us}$,  $\absV{cd}$. We then introduced 
a convenient parametrization of NP and examined the impact of 
$\Delta M_{B_{d}}$ and  $a_{J/\Psi K_S}$ in obtaining 
experimental evidence of a complex CKM matrix.
We then emphasized that the best 
evidence for CKM to be complex, arises from $\gamma$, 
either through a direct measurement or through a measurement of 
$\overline \alpha$, together with $\overline \beta$. 
We conclude that if NP does not pollute SM amplitudes dominated by tree level diagrams, a real CKM is excluded at a 99.92\% C.L., and the probability of having $\gamma$ in the region $[10^\circ;170^\circ]\cup[-170^\circ;-10^\circ]$ is 99.7\%.
We also illustrated how the above measurements can be used to 
place limits on the size of NP, which is allowed by the 
present data. As we emphasized in the Introduction,
having an irrefutable piece of evidence for a complex CKM matrix,
in a framework where the presence of NP is allowed, has 
profound implications for models of CP violation.
In the particular case of models with spontaneous CP
violation, a complex CKM matrix favours the class of models 
\cite{Branco:1985aq}, \cite{Bento:1990wv} 
 where, although Yukawa couplings
are real, the vacuum phase responsible for spontaneous 
CP violation also generates CP violation in charged-current weak
interactions. Conversely, the evidence for a complex CKM matrix,
even allowing for the presence of NP, 
excludes the class of models with spontaneous CP violation
and a real CKM matrix at a 99.92\% C.L..

\section*{Acknowledgements}
The authors thank F. Mart\'inez, J. Silva and Z. Ligeti for useful comments. 
GCB and MNR thank the CERN Physics Department (PH) Theory (TH) for the warm
hospitality. Their work was partially supported by CERN, by Funda\c c\~ao
para a Ci\^encia e a Tecnologia (FCT) (Portugal) through the Projects
PDCT/FP/FNU/ 50250/2003, PDCT/FP/FAT/50167/2003 and 
CFTP-FCT UNIT 777, which are partially funded through POCTI (FEDER)-
and also by Ac\c c\~ao Integrada Luso Espanhola E-73/04. 
MNR and GCB thank the
Universitat de Val\` encia for their very friendly welcome during 
productive working visits. 
FJB acknowledges the warm hospitality extended to him during his 
stay at IST, Lisbon, where part of this work was done. 
MN acknowledges the Spanish MEC for a fellowship. The work 
of FJB and MN was 
partially supported by MEC projects FPA2002-00612 and HP2003-0079.

\vspace{1cm}
\begin{center}
\begin{table}[bht]
\begin{tabular}{|c|c||c|c|}
\hline
$\absV{ud}$& $0.9738\pm 0.0005$ & $\absV{us}$& $0.2200\pm 0.0026$\\ \hline
$\absV{cd}$& $0.224\pm 0.012$ & $\absV{cs}$& $0.97\pm 0.11$\\ \hline
$\absV{ub}$& $(3.67\pm 0.47)10^{-3}$& $\absV{cb}$&
$(4.13\pm0.15)10^{-2}$\\ \hline
 $a_{J/\Psi K_{S}}$&$0.734\pm 0.054$&$A_{SL}$ &$(-3\pm 7)10^{-3}$ \\ \hline
$\gamma$ &$(69\pm 21)^\circ \mod 180^\circ$ &$\overline \alpha$ &
$(100\pm 16)^\circ \mod 180^\circ$\\ \hline
$\absV{us}_{Fig\ref{Fig:03}}$& $0.22000\pm 0.00002$ & &\\\hline
$\absV{cd}_{Fig\ref{Fig:03a}}$& $0.21985\pm 0.00002$ & $\absV{cd}_{Fig\ref{Fig:03b}}$& $0.21955\pm 0.00002$ \\ \hline\hline
$\Delta M_{B_{d}}$&$(0.502\pm 0.007)$ps$^{-1}$ & $m_{B_d}$& $5.2794$ GeV
\\ \hline
$f_{B_d}$&$(0.20\pm 0.03)$ GeV & $B_{B_d}$ & $(1.30\pm 0.18)$\\ \hline
$\eta_B$ & $0.55$ & $S_0(x_t)$ & $2.5745$\\ \hline
\end{tabular}
\caption{Numerical inputs for the different calculations \cite{PDBook,Group:2004cx}.}\label{TABLE}
\end{table}
\end{center}

\appendix
\section*{Appendix}

In this appendix we explain why the extraction of $\overline\alpha$ based in the so-called isospin analysis is valid even in the presence of New Physics in the $\Delta I=1/2$ piece of the $b\to d$ hamiltonian. This is true for $B\to\pi\pi$, $B\to\rho\rho$ and $B\to\rho\pi$. 

In the $B\to\pi\pi$ case, as has been rephrased in \cite{BotellaSilva}, the basic ingredients to extract $\overline\alpha$ are:
\begin{itemize}
\item[(i)\label{item:i}] The full hamiltonian only contains $\Delta I=1/2,3/2$ pieces, and isospin is a good symmetry of final state interactions. With these ingredients one has
\begin{equation}
\lambda_{+0}\equiv\frac{q}{p}\frac{\overline A_{+0}}{A_{+0}}=e^{-i2\overline\beta}\frac{\overline A_{3/2}}{A_{3/2}}~,\label{lambdapluszero}
\end{equation}
and $\lambda_{+0}$ is a ``physical observable'' that can be reconstructed (up to discrete ambiguities) with the directly measurable observables $\lambda_{+-}=\frac{q}{p}\frac{\overline A_{+-}}{A_{+-}}$ and the branching ratios of the 6 available channels $B^{i+j}\to\pi^i\pi^j$ and their CP-conjugates; obviously $A_{+0}=A(B^+\to\pi^+\pi^0)\propto A_{3/2}$ and $A_{+-}=A(B^0\to\pi^+\pi^-)$.
\item[(ii)] If the $\Delta I=3/2$ piece of the hamiltonian is exactly the SM one (the tree level piece \footnote{Inclusion of EWP gives a shift of $1.5^\circ$ in $\overline\alpha$ that we will neglect, although it can be included.}) we have
\begin{equation}
\lambda_{+0}=e^{-i2\overline\beta}e^{-i2\gamma}=e^{+i2\overline\alpha}~.\label{lambdapluszero:2}
\end{equation}
\end{itemize}
It is important to stress that the extraction of $\lambda_{+0}$ is valid in any model of NP that fulfills assumption (i), therefore the extraction of $\overline\alpha$ from equation (\ref{lambdapluszero:2}) is valid in any model where the NP in the decay amplitudes only appears in the $\Delta I=1/2$ piece. If full isospin analysis is not done, because the usual bounds are obtained assuming (i), the result is also valid. The $B\to\rho\rho$ case is similar to the $B\to\pi\pi$ case. 

In the Dalitz plot analysis of $B\to\rho\pi$, the moduli and the relative phases of $A^{+-}=A(B^0\to\rho^+\pi^-)$, $A^{-+}=A(B^0\to\rho^-\pi^+)$ and $A^{00}=A(B^0\to\rho^0\pi^0)$, are measured, together with the CP-conjugate channels, and a global relative phase weighted by $q/p$. With a general weak hamiltonian $H_w$ with $\Delta I=1/2,3/2$ pieces one obtains from isospin
\begin{equation}
e^{-i2\overline\beta}\frac{\overline A_{3/2,2}}{A_{3/2,2}}=e^{-i2\overline\beta}\frac{\overline A^{+-}+\overline A^{-+}+2\overline A^{00}}{A^{+-}+A^{-+}+2A^{00}}~,
\end{equation}
where $A_{3/2,2}$ is the reduced matrix element of $H_w(\Delta I=3/2)$ with a $I=2$ final state. Assuming that $H_w(\Delta I=3/2)$ is the SM one, we have
\begin{equation}
e^{-i2\overline\beta}\frac{\overline A_{3/2,2}}{A_{3/2,2}}=e^{-i2(\overline\beta+\gamma)}=e^{i2\overline\alpha}~.
\end{equation}
So the extraction of $\overline\alpha$ from the Dalitz plot analysis in $B\to\rho\pi$ is completely valid in the presence of any NP in the $\Delta I=1/2$ hamiltonian.


\newpage

\begin{thebibliography}{90}

\bibitem{Cabibbo:1963yz}
N.~Cabibbo,
\newblock Phys. Rev. Lett. {\bf 10}, 531 (1963).

M.~Kobayashi and T.~Maskawa,
\newblock Prog. Theor. Phys. {\bf 49}, 652 (1973).

\bibitem{Bona:2005vz}
UTfit, M.~Bona {\em et~al.},
\newblock  hep-ph/0501199, \\ \texttt{http://www.utfit.org}


CKMfitter Group, J.~Charles {\em et~al.},
\newblock   hep-ph/0406184, \\ \texttt{http://ckmfitter.in2p3.fr}


A.~J. Buras, F.~Parodi, and A.~Stocchi,
\newblock JHEP {\bf 01}, 029 (2003), hep-ph/0207101.

\bibitem{Branco:1999sx}
G.~C. Branco, F.~Cagarrinho, and F.~Kruger,
\newblock Phys. Lett. {\bf B459}, 224 (1999), hep-ph/9904379.

\bibitem{Lee:1973iz}
T.~D. Lee,
\newblock Phys. Rev. {\bf D8}, 1226 (1973).

\bibitem{Branco:1985aq}
G.~C. Branco and M.~N. Rebelo,
\newblock Phys. Lett. {\bf B160}, 117 (1985).

J.~Liu and L.~Wolfenstein,
\newblock Nucl. Phys. {\bf B289}, 1 (1987).


\bibitem{Bento:1990wv}
L.~Bento and G.~C. Branco,
\newblock Phys. Lett. {\bf B245}, 599 (1990).

L.~Bento, G.~C. Branco, and P.~A. Parada,
\newblock Phys. Lett. {\bf B267}, 95 (1991).

\bibitem{Branco:2000dq}
G.~C. Branco, F.~Kruger, J.~C. Romao, and A.~M. Teixeira,
\newblock JHEP {\bf 07}, 027 (2001), hep-ph/0012318.

C.~Hugonie, J.~C. Romao, and  A.~M. Teixeira,
\newblock JHEP {\bf 06}, 020 (2003), hep-ph/0304116.

\bibitem{Branco:1979pv}
G.~C. Branco,
\newblock Phys. Rev. Lett. {\bf 44}, 504 (1980).

G.~C. Branco,
\newblock Phys. Rev. {\bf D22}, 2901 (1980).

G.~C. Branco, A.~J. Buras, and J.~M. Gerard,
\newblock Nucl. Phys. {\bf B259}, 306 (1985).

\bibitem{Aubert:2004qm}
BaBar, B.~Aubert {\em et~al.},
\newblock Phys. Rev. Lett. {\bf 93}, 131801 (2004), hep-ex/0407057.

Belle, Y.~Chao {\em et~al.},
\newblock Phys. Rev. Lett. {\bf 93}, 191802 (2004), hep-ex/0408100.

\bibitem{Branco:1999fs}
G.~C.~Branco, L.~Lavoura and J.~P.~Silva, ``CP
violation,'' International Series of Monographs on Physics, No. 103, Oxford
University Press. Oxford, UK: Clarendon (1999), 511 pp.

\bibitem{Aleksan:1994if}
R.~Aleksan, B.~Kayser, and D.~London,
\newblock Phys. Rev. Lett. {\bf 73}, 18 (1994), hep-ph/9403341.

\bibitem{Botella:2002fr}
F.~J. Botella, G.~C. Branco, M.~Nebot, and M.~N. Rebelo,
\newblock Nucl. Phys. {\bf B651}, 174 (2003), hep-ph/0206133.

J.~A. Aguilar-Saavedra, F.~J. Botella, G.~C. Branco, and M.~Nebot,
\newblock Nucl. Phys. {\bf B706}, 204 (2005), hep-ph/0406151.

\bibitem{SilvaWolf}
J.~Silva, and L.~Wolfenstein,
\newblock Phys. Rev. {\bf D55}, 5331 (1997), hep-ph/9610208.

\bibitem{Charles:2004jd}
CKMfitter Group, J.~Charles {\em et~al.}, same as in \cite{Bona:2005vz}.

\bibitem{Ligeti:2004ak}
Z.~Ligeti,
\newblock  hep-ph/0408267.

\bibitem{Botella:1985gb}
F.~J. Botella and L.-L. Chau,
\newblock Phys. Lett. {\bf B168}, 97 (1986).

G.~C. Branco and L.~Lavoura,
\newblock Phys. Lett. {\bf B208}, 123 (1988).

\bibitem{Jarlskog:1985ht}
C.~Jarlskog,
\newblock Phys. Rev. Lett. {\bf 55}, 1039 (1985).

\bibitem{Cowan:1998ji}
G.~Cowan,
\newblock Oxford, UK: Clarendon, (1998), 197 pp.

G.~D'Agostini,
\newblock CERN Report 99-03 (1999).

G.~D'Agostini,
\newblock Rep. Prog. Phys. {\bf 66}, 1383 (2003), physics/0304102.

\bibitem{Randhawa:2001de}
M.~Randhawa and M.~Gupta,
\newblock  hep-ph/0102274.

M.~Randhawa and M.~Gupta,
\newblock Phys. Lett. {\bf B516}, 446 (2001), hep-ph/0106161.

P.~Dita,
\newblock  hep-ph/0408013.

\bibitem{Chau:1984fp}
L.-L. Chau and W.-Y. Keung,
\newblock Phys. Rev. Lett. {\bf 53}, 1802 (1984).

\bibitem{PDBook}
S.~Eidelman {\it et al.}  [Particle Data Group Collaboration],
Phys.\ Lett.\ B {\bf 592}, 1 (2004)

\bibitem{Atwood:2003tg}
D.~Atwood and G.~Hiller,
\newblock  hep-ph/0307251.

\bibitem{Gronau:1990ra}
M.~Gronau and D.~London,
\newblock Phys. Lett. {\bf B253}, 483 (1991).

M.~Gronau and D.~Wyler,
\newblock Phys. Lett. {\bf B265}, 172 (1991).

I.~Dunietz,
\newblock Phys. Lett. {\bf B270}, 75 (1991).

R.~Aleksan, I.~Dunietz, and B.~Kayser,
\newblock Z. Phys. {\bf C54}, 653 (1992).

D.~Atwood, G.~Eilam, M.~Gronau, and A.~Soni,
\newblock Phys. Lett. {\bf B341}, 372 (1995), hep-ph/9409229.

D.~Atwood, I.~Dunietz, and A.~Soni,
\newblock Phys. Rev. Lett. {\bf 78}, 3257 (1997), hep-ph/9612433.

A.~Bondar and T.~Gershon,
\newblock Phys. Rev. {\bf D70}, 091503 (2004), hep-ph/0409281.

\bibitem{Giri:2003ty}
A.~Giri, Y.~Grossman, A.~Soffer, and J.~Zupan,
\newblock Phys. Rev. {\bf D68}, 054018 (2003), hep-ph/0303187.

\bibitem{Gronau:1990ka}
M.~Gronau and D.~London,
\newblock Phys. Rev. Lett. {\bf 65}, 3381 (1990).

Y.~Grossman and H.~R. Quinn,
\newblock Phys. Rev. {\bf D58}, 017504 (1998), hep-ph/9712306.

J.~Charles,
\newblock Phys. Rev. {\bf D59}, 054007 (1999), hep-ph/9806468.

M.~Gronau, D.~London, N.~Sinha, and R.~Sinha,
\newblock Phys. Lett. {\bf B514}, 315 (2001), hep-ph/0105308.

\bibitem{Snyder:1993mx}
A.~E. Snyder and H.~R. Quinn,
\newblock Phys. Rev. {\bf D48}, 2139 (1993).

\bibitem{Silva:2004gz}
J.~P. Silva,
\newblock  hep-ph/0410351.

\bibitem{Barenboim:1997qx}
G.~Barenboim, F.~J. Botella, G.~C. Branco, and O.~Vives,
\newblock Phys. Lett. {\bf B422}, 277 (1998), hep-ph/9709369.

\bibitem{Abe:2004gu}
Belle, K.~Abe {\em et~al.},
\newblock  hep-ex/0411049.

\bibitem{Aubert:2004kv}
BaBar, B.~Aubert {\em et~al.},
\newblock  hep-ex/0408088.

\bibitem{Amorim:1998pi}
A.~Amorim, M.~G. Santos, and J.~P. Silva,
\newblock Phys. Rev. {\bf D59}, 056001 (1999), hep-ph/9807364.

C.~C. Meca and J.~P. Silva,
\newblock Phys. Rev. Lett. {\bf 81}, 1377 (1998), hep-ph/9807320.

J.~P. Silva and A.~Soffer,
\newblock Phys. Rev. {\bf D61}, 112001 (2000), hep-ph/9912242.

\bibitem{BabarSignCos2Beta:2004}
BaBar, B.~Aubert {\em et~al.},
\newblock  hep-ex/0411016.

Belle, K.~Abe {\em et~al.},
\newblock  hep-ex/0408104.

\bibitem{Aubert:2004pt}
BaBar, B.~Aubert,
\newblock  hep-ex/0408038.

\bibitem{fernando}
Private communication, Fernando Mart\'inez Vidal.

\bibitem{0502139Gronau}
M. Gronau and J. Zupan,
\newblock Phys. Rev. {\bf D71}, 074017 (2005), hep-ph/0502139.

\bibitem{Aubert:2004zr}
BaBar, B.~Aubert {\em et~al.},
\newblock Phys. Rev. Lett. {\bf 93}, 231801 (2004), hep-ex/0404029.

\bibitem{Aubert:2003wr}
BaBar, B.~Aubert {\em et~al.},
\newblock Phys. Rev. Lett. {\bf 91}, 201802 (2003), hep-ex/0306030.

\bibitem{Zhang:2003up}
Belle, J.~Zhang {\em et~al.},
\newblock Phys. Rev. Lett. {\bf 91}, 221801 (2003), hep-ex/0306007.

\bibitem{Bevan:2004ht}
BaBar, A.~Bevan,
\newblock  hep-ex/0411090.

\bibitem{Aubert:2004iu}
BaBar, B.~Aubert {\em et~al.},
\newblock  hep-ex/0408099.

\bibitem{Beneke:2003az}
M.~Beneke, G.~Buchalla, A.~Lenz, and U.~Nierste,
\newblock Phys. Lett. {\bf B576}, 173 (2003), hep-ph/0307344.

M.~Ciuchini, E.~Franco, V.~Lubicz, F.~Mescia, and C.~Tarantino,
\newblock JHEP {\bf 08}, 031 (2003), hep-ph/0308029.

\bibitem{Branco:1992uy}
G.~C. Branco, P.~A. Parada, T.~Morozumi, and M.~N. Rebelo,
\newblock Phys. Lett. {\bf B306}, 398 (1993).

S.~Laplace, Z.~Ligeti, Y.~Nir, and G.~Perez,
\newblock Phys. Rev. {\bf D65}, 094040 (2002), hep-ph/0202010.


\bibitem{BotellaSilva}
F.J. Botella and J. Silva,
\newblock Phys. Rev. {\bf D71}, 094008 (2005), hep-ph/0503136.

\bibitem{Group:2004cx}
Heavy Flavour Averaging Group
\newblock , hep-ex/0412073.

\newblock \texttt{http://www.slac.stanford.edu/xorg/hfag/}

\end{thebibliography}

\end{document}